\documentclass[12pt,preprint]{aastex}

\begin{document}

\title{Radial Gas Flows in Colliding Galaxies: Connecting Simulations and Observations}

\author {Daisuke Iono\altaffilmark{1,2}, Min S. Yun\altaffilmark{1}, J. Christopher  Mihos\altaffilmark{3,4}}
\altaffiltext{1}{Department of Astronomy, University of Massachusetts, Amherst, MA 01002} 
\altaffiltext{2}{Harvard-Smithsonian Center for Astrophysics, 60 Garden Street, Cambridge, MA 02138}
\altaffiltext{3}{Department of Astronomy, Case Western Reserve University, Cleveland, OH 44106}
\altaffiltext{4}{Research Corporation Cottrell Scholar and NSF CAREER Fellow}

\begin{abstract}

We investigate the detailed response of gas to the formation of
transient and long-lived dynamical structures induced in the early
stages of a disk-disk collision, and identify observational signatures
of radial gas inflow through a detailed examination of the collision
simulation of an equal mass bulge dominated galaxy.
Our analysis and discussion mainly focuses around the evolution of the
diffuse and dense gas in the early stages of the collision, when the
two disks are interacting but have not yet merged.  Stars respond to
the tidal interaction by forming both transient arms and long lived
$m=2$ bars, but the gas response is more transient, flowing directly
toward the central regions within about $10^8$ years after the initial
collision.  The rate of inflow declines when more than half of the
total gas supply reaches the inner few kpc, where the gas forms a
dense nuclear ring inside the stellar bar.  The average gas inflow
rate to the central 1.8 kpc is $\sim 7 M_\odot$ yr$^{-1}$ with a peak
rate of $17 M_\odot$ yr$^{-1}$.  Gas with high volume density is found
in the inner parts of the post-collision disks at size scales close to
the spatial resolution of the simulations, and this may be a direct
result of shocks traced by the discontinuity in the gas velocity field.
The evolution of gas in a bulgeless progenitor galaxy is also
discussed, and a possible link to the ``chain galaxy'' population
observed at high redshifts is inferred.  The evolution of the
structural parameters such as asymmetry and concentration of both
stars and gas are studied in detail.  Further, a new structure
parameter (the compactness parameter $K$) that traces the evolution of
the size scale of the gas relative to the stellar disk is introduced,
and this may be a useful tracer to determine the merger chronology of
colliding systems.  Non-circular gas kinematics driven by the
perturbation of the non-axisymmetric structure can produce distinct
emission features in the "forbidden velocity quadrants'' of the
position-velocity diagram (PVD).  The dynamical mass calculated using
the rotation curve derived from fitting the emission envelope of the
PVD can determine the true mass to within 20\% to 40\%.  The evolution
of the molecular fraction ($M_{\rm H_2}/M_{\rm H_2 + HI}$) is a
potential tracer to quantitatively assign the age of the interaction,
but the application to real systems may require additional
observational diagnostics to properly assess the exact chronology of
the merger evolution.

\end{abstract}

\keywords{galaxies: interactions, kinematics and dynamics --- methods: numerical --- galaxies: individual : UGC~12914/15 Taffy}

\section{Introduction}

Collisionally-induced radial inflow of gas is one of the primary
mechanisms that forces significant amount of gas to the central kpc of
galaxies.  Such inward streaming in gas-rich progenitor galaxies can
trigger the onset of intense nuclear bursts of star formation,
resulting in systems large in infrared luminosity
\citep[i.e. ULIRGs;][]{sanders96, genzel00}, or supply the fuel for
central AGN activity \citep{kennicutt84}, or possibly both.  Recent
high resolution imaging by space-borne observatories found that the
majority of the ULIRGs display highly disturbed optical and infrared
morphology that signify recent or ongoing strong tidal activity
\citep{farrah01,bush02}.  While some of the ULIRGs are believed to be
in the intermediate stages of merging based on their projected nuclear
separation (some as large as 20 kpc), the majority are in a more
advanced stage similar to the prototypical nearby ULIRG Arp 220, where
the projected nuclear separation of $\sim 0.3$ kpc is found from high
resolution radio synthesis images \citep{sakamoto99}.  Such merger
events are important not only to understand the underlying physics
that governs galaxy dynamics in the local universe, but are also key
events in the formation and evolution of galaxies in the early
universe, where the mean distance between galaxies was much smaller at
present.  Based on a number of observational similarities, these local
ULIRGs are now commonly invoked as the closest analogs of high
redshift sub-mm sources \citep{blain02}.

It is widely believed that the perturbation of the stellar $m=2$ mode
(i.e. bars and arms) is the key mechanism in initiating the massive
radial gas inflow.  Disk instability in isolated systems
\citep{kalnajs71, kalnajs72, miller78, combes81, sellwood81,
weinberg85} and forced perturbation in colliding systems \citep[][MH96
hereafter]{barnes96,mihos96} are mostly responsible for the growth of
non-axisymmetric modes.  To complement the theory, statistical
evidence suggests a higher fraction ($\sim$ 50\%) of barred galaxies
among interacting systems than among isolated spirals
\citep{elmegreen90}.  There is also evidence of enhanced fraction of
barred galaxies in the central region of the Coma cluster, where tidal
influences are more effective than in the outskirts of the cluster
\citep{thompson81}.  Results from steady-state hydrodynamical
simulations in such barred potentials emphasize the importance of the
orbital resonance (Inner/Outer Lindblad Resonance:ILR/OLR) in
promoting both central gas inflow
\citep[e.g.][]{athanassoula92b,sellwood93,byrd94} and radial outflow
\citep{buta96}.  These results further argue that the crossing of the
orthogonal orbits ($x_1$ \& $x_2$ orbits according to the nomenclature
of \citet{contopoulos80}) is the main mechanism that forces
dissipation of energy.

Some of the early numerical studies of colliding systems demonstrate
the importance of forced perturbation that can disturb the stellar
disk with enough amplitude to initiate the radial inflow
\citep[e.g.][]{noguchi88}.  In more recent studies, development of the
bar-like potential and resulting inflow of more than half of the gas
into the central few hundred parsecs of the merger remnant has been
demonstrated \citep[][MH96]{barnes96}.  MH96 found in their
investigation that major merger events trigger violent central
streaming of gas followed by significant starburst activity, the
intensity of which could account for the infrared luminosity of the
UILRGs.  They found, by isolating the torque imposed on the infalling
gas particles, that the primary mechanism responsible for the radial
gas inflow is $gravitational$ torque induced primarily from the stars
in the host galaxy.  They allude to the importance of hydrodynamical
processes in initiating this gas inflow, but did not explore the
evolution of the gas distribution on small scales in any detail.  In
addition, MH96 found that the detailed dynamical structure such as the
presence/absence of a dense stellar bulge can strongly influence the
intensity and the triggering of the star formation activity. For
progenitor galaxies with central bulge support, mild star formation
activity occurs continuously shortly after the initial contact, while
the final merger of the nuclei is characterized by a more vigorous and
acute star formation activity.  On the other hand, moderate degree of
star formation activity occurs for over a longer time period ($\sim 2
\times 10^8$ years) in the initial stages of the bulgeless galaxy
encounter.  Much of the gas in the central region of the bulgeless
galaxies is depleted and converted into stars before the final merger.
In both cases, a total of nearly $75\%$ of the initial gas has been
converted to stars (modeled using a density-dependent Schmidt law)
throughout the entire merger sequence.  Finally, they investigated the
effects of varying encounter geometry and concluded that the
dependence on the orbital parameters is less significant than the
non-axisymmetric response of the host galaxy (i.e. bars and arms) in
initiating the inflow activity.

Because the main intent of the paper by MH96 was to investigate the 
development of gas inflow and the degree of starbursts that follow, a 
detailed analysis of the morphological and dynamical evolution of the 
stars and gas with each colliding disks were not discussed.   Furthermore, 
self consistent analysis of numerical models can offer valuable insight 
to the observational interpretation of the distribution and kinematics 
of the ISM in colliding disk systems. Therefore, to further understand 
and extend the numerical analysis that were beyond the scope in their 
original analysis in MH96, and to develop a set of analysis tools to 
help understand the observational data better, a thorough reanalysis 
of their numerical experiment in MH96 is conducted.  Some of the 
analysis include for example; (1) the detailed response of the gas 
under a varying stellar potential, (2) the characteristics of the 
inflow and its consequences, (3) the sensitivity of the gas response 
to the underlying stellar structure, (4) the evolution of observable 
parameters such as asymmetry and concentration, (5) the observational 
interpretation of the inflow, and (6) the determination of the interaction 
chronology of merging pairs of galaxies through the use of observable 
parameters.  Because the intent of this study is to characterize the 
development of axi-symmetric structure in stars and the subsequent 
inflow activity in gas, the analysis will focus on the disk evolution 
mostly up to the pre-coalescence phase when the disks are interacting 
but have not yet merged. It will be shown that the resulting inflow 
period lasts for $\sim 10^8$ years (compare with the merger time scale 
of $\sim 5 \times 10^8$ to $10^9$ years), suggesting that $10 - 20\%$ 
of randomly selected observed pairs of interacting galaxies may show 
evidence of such radial inflow.  In paper II, the observational tools 
developed here will be tested toward \ion{H}{1} and CO~(1--0) observations 
of 10 early stage interacting systems, and paper III discusses the results 
from a detailed comparison with a control sample that includes isolated 
galaxies from the BIMA SONG survey \citep{helfer03}.

The numerical model and the units adopted are briefly summarized in \S2.  
Detailed analysis of the gas response to the stellar $m=2$ features, the 
inflow mechanism, mass inflow rate, structural evolution of gas, and some 
discussion on the role of shocks can be found in \S3 and \S4.  In \S5, 
the observational implication from the gas kinematics is discussed, and 
the possible use of molecular fraction ($M_{H_2}/M_{H_2 + HI}$) is 
demonstrated in an attempt to infer the merger chronology in \S6.

\section{Numerical Model}

The adopted TREESPH technique \citep{hernquist89} is a hybrid between
an n-body evolution code developed by arranging the particles in an
oct-tree structure \citep{barnes86}, and the smoothed particle
hydrodynamics (SPH) method that allows a gird-free computation of the
evolution of the gas and its properties \citep{lucy77, gingold77}.
The initial distribution of the progenitor galaxies are completely
identical, but differs in their geometry of the orbits; disk 1 is
placed in a prograde geometry whereas disk 2 is highly inclined with
respect to the orbital plane.  The galaxies are equal in mass,
constructed with a stellar and gaseous disk, a stellar bulge with a
bulge-to-disk mass ratio of 1/3, and a dark halo (MH96).  Each SPH
particle has two mass components; the total mass which is used to
calculate the gravitational evolution, and the gas mass which is used
to calculate the hydrodynamical forces and the associated properties
of the gas \citep{mihos94}. The rate of conversion from gas to stars
is determined by the smoothed local gas density, where the hybrid SPH
particle is converted to a young star when the gas mass fraction falls
below 5\% of the total \citep{mihos94}.  The ISM is modeled with an
isothermal equation of state which fixes the gas temperature at $10^4$
K, despite the evidence that the real ISM harbors gas in multiple
temperature \citep{mckee77}.  Such isothermal treatment is often
justified by the short cooling time compared to the time resolution
adopted for the simulations.  Even in simulations that explicitly
calculate heating and cooling above and below $10^4$ K, the shock
heated gas immediately radiates away its energy resulting in most of
the gas staying near the adopted value of $10^4$ K since the cooling
curve plummets below this temperature \citep{hernq89, barnes91,
mihos96, barnes01}.  If the simulation allowed cooling below $10^4$ K,
clumps of high density gas clouds could form, which might represent
clusters of active star formation in real systems.  The isothermal
approximation can yield incorrect results if the goal is to study the
detailed kinematics of the ISM at less then the kpc scale
\citep{barnes01}. However, since our aim is to characterize the large
scale gas streaming motion at scales comparable to or larger than a
kpc, the less computationally expensive, isothermal simulation should
serve our purpose.

MH96 studied 8 independent simulation results that are different in
the disk structure and the orbital geometry of the encounter. The four
different orbital geometries include; prograde-inclined,
prograde-prograde, prograde-retrograde, and inclined-inclined.  For
each encounter geometry, two different types of initial conditions
were adopted for the disk density structure, one of which has a bulge
component dominating the central potential (bulge dominated galaxy)
and the other without the central support (bulgeless galaxy),
comprising a total of eight different models.  The most realistic case
in the local universe -- the encounter of two bulge dominated galaxies
on prograde-inclined orbit -- is mainly examined throughout this
study.  Some quantitative analysis of the experiments using bulgeless
models are also discussed, where such a scenario may be more
representative of the early universe if the progenitor galaxies were
more disk-dominated.

Unless otherwise noted, all simulation units are scaled to those of
the Milky Way galaxy, which gives unit length of 3.5 kpc, unit mass of
$5.6 \times 10^{10}$M$_{\odot}$, unit velocity of 262 km/s, and unit
time of $\sim 1.3 \times 10^7$ years (MH96).  The time step used in
the simulation is $1.0 \times 10^6$ years.  The above normalization
scales each of the equal mass gas particles to $3 \times
10^5$M$_\odot$, mass comparable to a typical GMC in the Milky Way
galaxy.  Since the main intent is to study the dynamical evolution of
the disks after the initial collision, the pericentric passage is
defined as $t=0$ and time advances forward thereafter, while the epoch
before the collision is represented by negative time.  The above
normalization and all references to physical scales should be treated
as illustrative because the derived physical properties could vary
according to the initial conditions of the host galaxies.  For
instance, the characteristic size and the extent of the bar length is
sensitive to the orbital properties and the bulge-to-disk ratio of the
host galaxy \citep{combes93}.  Thus, in what follows, the physical
units are used only to offer a sense of the size and time scales in a
sensible and convenient manner.

\section{Results}
\subsection{Formation and Evolution of the Stellar Structures and the Associated Gas Response}

The merger sequence is presented in Figure~\ref{fig1} along with a
side-by-side comparison of stars and gas in each disk viewed face-on.
Each SPH particle is smoothed by its individual smoothing length in
order to display the gas as a continuous distribution, and the
overlaid contours represent the logarithmic density distribution of
stars.  Similar but more detailed evolution of the gas in disk 1 is
presented in Figure~\ref{fig2}.  Here, the distribution of gas is
color coded according to the smoothed local volume density, where
blue, green, yellow, and red each represent densities above $n_{\rm
gas}$ = 0.1, 1, 10 and $100$ cm$^{-3}$ averaged over their smoothing
lengths respectively.  Also plotted in the upper left side of each
panel is the approximate orientation of the major-axis of the stellar
bar.  Only the evolution of disk 1 is presented here -- the
distribution of gas in the inner regions of disk 2 undergoes a similar
evolution to that in disk 1 (despite the different mass evolution seen
in the {\it outer} disk due to the difference in the encounter
geometry; see \S 3.4).  The local gas density governs the length of a
typical smoothing length that ranges from 50 pc to 5 kpc; a difference
of two orders of magnitude between the high density nuclear gas and
the more diffuse gas populating the outskirts of the galaxy.  Since
the vertical disk scale hight ($z_0 = 700$ pc) is comparable to or
less than the SPH smoothing lengths, structure in the vertical
direction is often unresolved \citep{barnes02}.  Therefore, only the
snapshots of the inner disk are provided in order to study the
detailed structure of gas resolved with at least a few hundred parsec
scale.  Inspection of the merger sequence in Figure~\ref{fig1} \& 2
reveals striking difference in both the stellar and gaseous morphology
before and after the collision, and presents several distinct
features; (1) growth of the large scale stellar $m=2$ features, (2)
subsequent gas response and (3) the formation of the nuclear gas ring.
Some of the key features seen in each evolutionary stages shown in
Figure~\ref{fig1} \& 2 are discussed in the following subsections.

\clearpage

\begin{figure*}
\epsscale{0.9}
\caption{The projected view of the collision is shown ($left$), along with 
the face on view of disk 1 ($middle$) and disk 2 ($right$).  The contours 
in the face-on plots represent the logarithmic stellar distribution, and 
the logarithmic gas distribution is shown in gray scale. Each frame is 
14kpc in both axes. Time advances from top to the bottom where each frame 
represents $-1.9 \times 10^8$, 0, $1.2 \times 10^8$ and $2.8 \times 10^8$ 
years.}
\label{fig1}
\end{figure*}

\begin{figure*}
\epsscale{1}
\caption{The evolution of the gas density distribution color coded
according to the local volume density.  The labels in each panel are
time in units of $10^7$ years, and the boxes are 14 kpc on a side.
Blue, green, yellow, and red represent $n_{\rm gas} \geq 0.1,1,10$ and
$100$ cm$^{-3}$, respectively.  The approximate orientation of the bar
is shown in the upper left corner of each snapshot for comparison with
the distribution of the gas.}
\label{fig2}
\end{figure*}

\clearpage

\subsubsection{Pre-Collision Disks ($t=-1.9 \times 10^8$ years)}

During the initial stages of the encounter, the stellar disks
generally maintain their initial distribution because the tidal
influence from the companion galaxy is negligible at this time.  On
the other hand, gas particles in both disks develop similar transient
spiral structure as seen in Figure~\ref{fig1} and 2.  Such phenomenon
-- swing amplification of particle noise -- is often observed in
n-body simulations with limited particle resolution, where gas is more
susceptible than stars to such disk instability because of its
dissipative nature \citep{barnes96}.

\subsubsection{The Massive Inflow Period ($t=0$ to $t \sim 1.0 \times 10^8$ years)}

The distribution of gas (see \S 3.4) shows significant evolution
during the first $10^8$ years after the collision when the strong
collisional perturbation at the pericentric passage ($\Delta_{disk}
\sim 9$ kpc) give rise to the excitation of the stellar bar and the
tidal arms ($m=2$ modes).  Tidal forces from the companion galaxy are
responsible for the perturbation of the non-axisymmetric features
\citep{toomre72}.  The growth of the stellar $m=2$ modes forces rapid
accumulation of gas into stellar arms that subsequently forms highly
shocked, dense filaments whose dissipational nature allows rapid
radial streaming of large quantities of disk gas.  The massive inflow
continues for $\lesssim 1.0 \times 10^8$ years until substantial
amount of gas supply in the periphery of the disks has either funneled
to the nuclear region or has been radially ejected into tidal tails.
Unlike the stellar arms which disperse within a few bar rotation
period, the induced stellar bar persists ubiquitously until the final
coalescence, with a rotation period roughly twice the time it takes
for the massive inflow to slow down (i.e. $P_{bar} \approx 2 \times
10^8$ years).

Comparing the modeled inflow timescale ($10^8$ years) with the total
merger timescale ($5 - 10 \times 10^8$ years) suggest that 10 - 20\%
of randomly selected strongly interacting/merging sytems in the local
universe might exhibit observable signature of radial inflow. This
number is based on the assumption that any observed sample of
interacting systems are comprised of systems which bear a resemblance
to the simulations, i.e., involve comparable mass galaxies
experiencing a close interaction, are past first pericentric passage,
and are made of galaxies which are structurally similar to the
simulated progenitors.  Variation in these properties will likely
modify the true fraction of galaxies with signature of inflow to some
degree.  This hypothesis will be tested in paper II by observing
atomic and molecular gas in ten interacting systems in the local
universe.

\begin{figure*}
\epsscale{1}
\plottwo{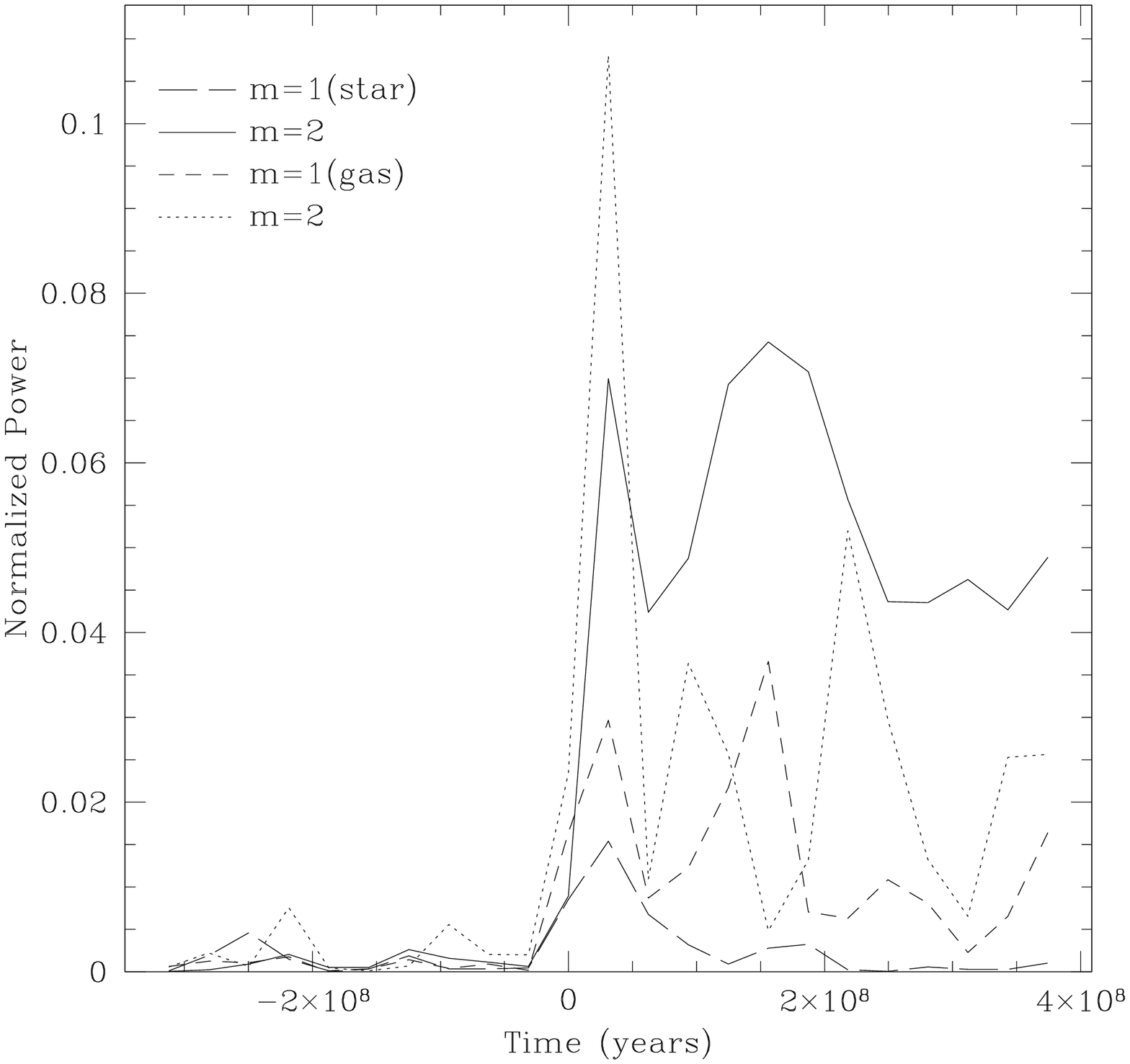}{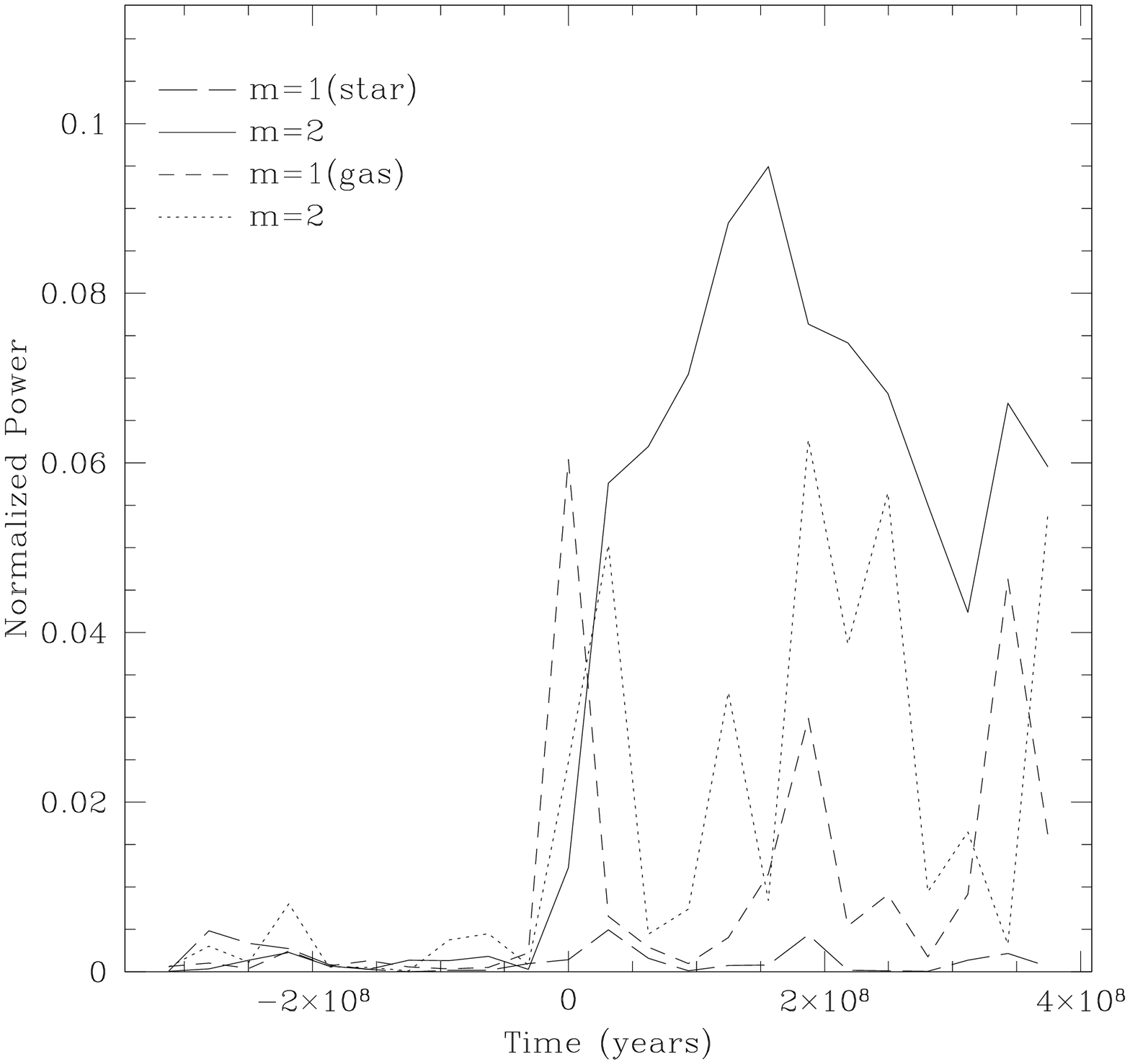}
\caption{The evolution of the m=1 and m=2  modes from a Fourier analysis 
of disk 1 ($left$) and disk 2 ($right$) using gas and star particles inside 
a sphere enclosed within $R=3.5$ kpc (see Figure 6 ($left$) for the spatial 
scale).  The Fourier amplitudes are normalized with the m=0 mode.}
\label{fig3}
\end{figure*}

\begin{figure*}
\epsscale{1}
\plottwo{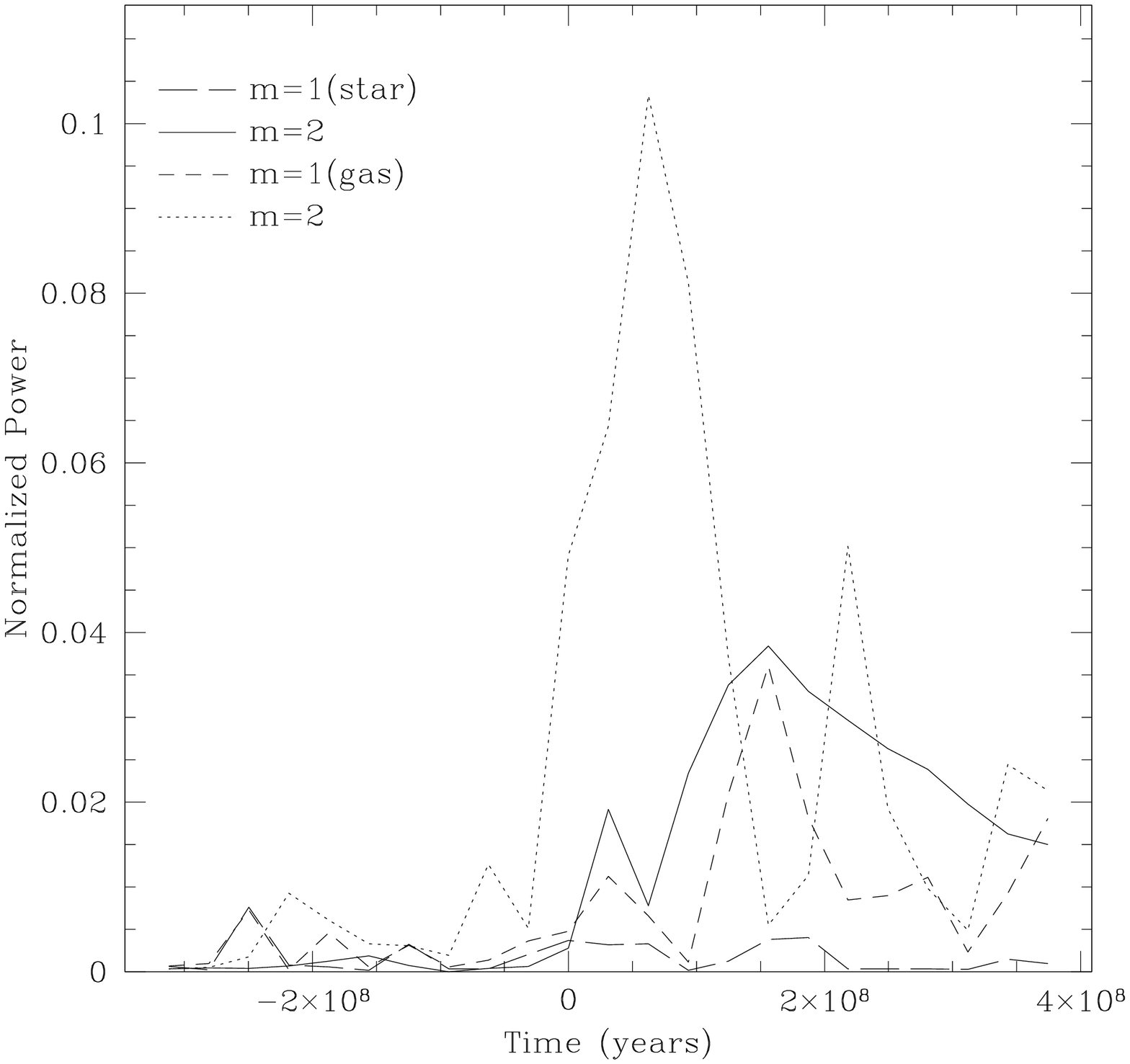}{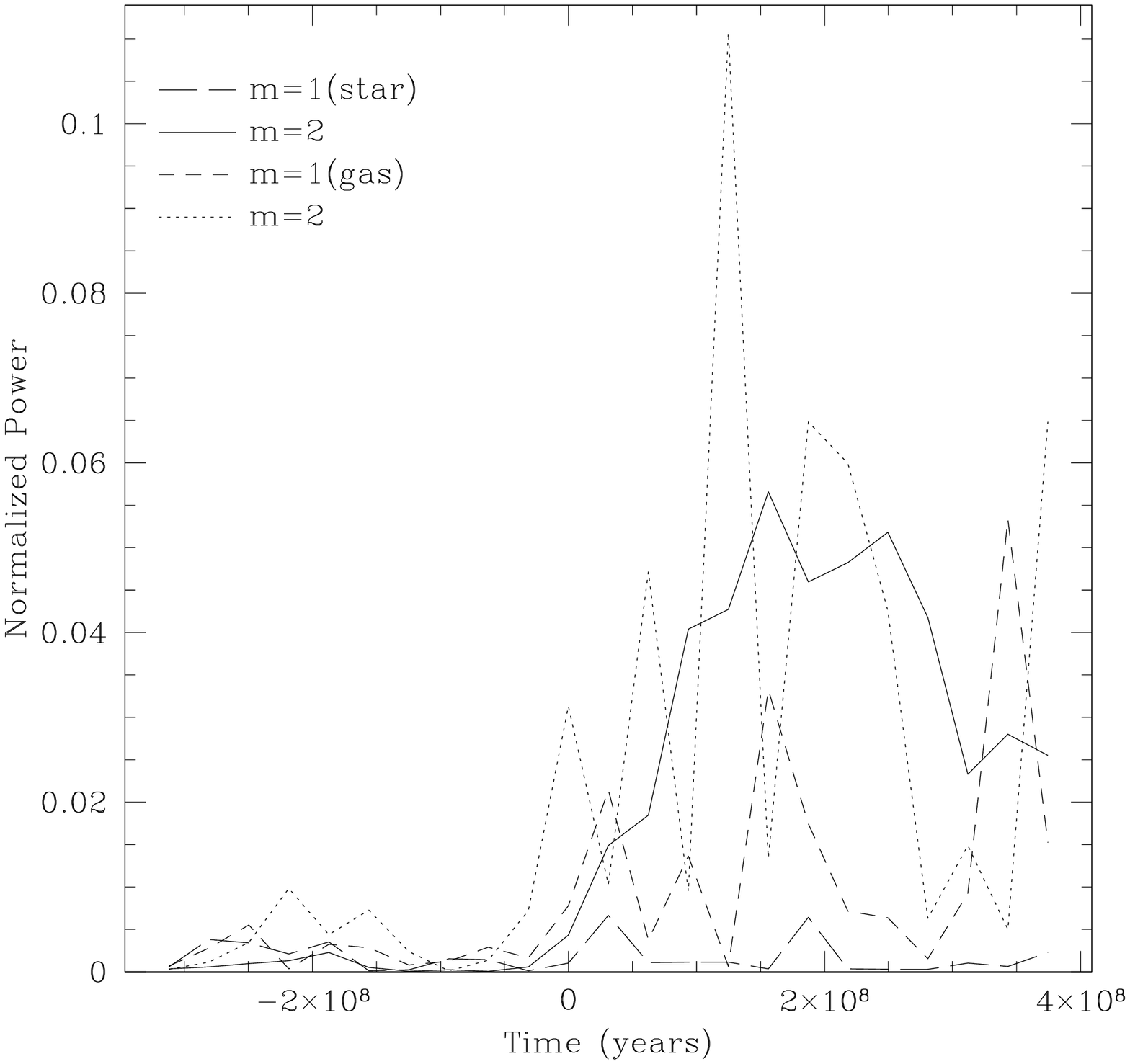}
\caption{Similar to Figure~\ref{fig3} but using particles inside a sphere 
enclosed within $R=1.8$kpc.}
\label{fig4}
\end{figure*}

A Fourier modal analysis allows a quantitative way to assess the
strength and duration of the excitation of the modes.  In order to
trace the modes forming inside and outside the radial extent of the
stellar bar, all of the star (including the disk and the bulge) and
gas particles are divided into equally spaced azimuthal bins with
radius restricted from $R = 0 - 3.5$ kpc and $R = 0 - 1.8$ kpc.  The
results for $R=0-3.5$ kpc are shown in Figure~\ref{fig3} for disk 1
($right$) and disk 2 ($left$), displaying here up to the dipole
component.  Similarly, the results for $R=0-1.8$ kpc are shown in
Figure~\ref{fig4}.  In general, the evolution of all modes are similar
between disk 1 and disk 2, but subtle differences arise due to the
transient nature of the modes in gas (see below).  The initial jump in
the $m=2$ mode in stars is attributed to the immediate formation of
the central bar whose ends are attached with the arms that are out of
phase by $180^\circ$.  This is more evident in $R = 0 - 3.5$ kpc
(Figure~\ref{fig3}) than in $R = 0 - 1.8$ kpc (Figure~\ref{fig4})
because the latter is inside the edge of the stellar bar and the bulge
dominates the dynamics in the inner disk.  At the same time, the gas
shows a stronger transient signal as it immediately responds to the
formation of the stellar $m=2$ structure.  The stronger $m=2$ signal
in gas arises in part due to the highly concentrated nature of gas,
but also due to the fact that the stellar calculation involves both
the disk and (less responsive) bulge stars.  While the stellar bar
dominates and maintains a strong $m=2$ signal in $R = 0 - 3.5$ kpc
after the massive inflow period and until coalescence (t $> 10^8$
years), the behavior of gas during this period is highly transient and
it is dominated by the distribution of gas at the central kpc scale.
Therefore, the evolution of $m=2$ and $m=1$ in gas are quite similar
after t $> 10^8$ years regardless of the radius (i.e the evolution of
$m=2$ and $m=1$ in gas in Figure~\ref{fig3} and Figure~\ref{fig4} are
similar after t $\sim 10^8$ years).

Spikes in the gaseous $m=1$ mode reach amplitudes slightly smaller
than that seen in the $m=2$ mode, but the transient nature of the
$m=1$ mode during (t $< 10^8$ years) and after (t $> 10^8$ years) the
massive inflow period is evident.  The stochastic nature of the
occurrence of the $m=1$ mode seen in simulations has an important
observational consequence.  Recent, detailed investigation of the
nuclear molecular gas in a few nearby galaxies with high spatial
resolution has revealed a distribution of molecular gas that is
lopsided toward one side of the galaxy \citep{garcia03, combes04}.  A
natural explanation is to assume a possible role of an $m=1$
perturbation that forces direct central transportation of gas which
may eventually become the source of the fuel for the central AGN
\citep{combes03}.  However, since the simulation results imply that
the $m=1$ mode is highly transient and could occur anytime throughout
the entire merger timescale, persistent central fueling through the
$m=1$ mode seems unlikely.  This interpretation is not limited to
major interacting systems as perturbation of a bar could occur
intrinsically or from a minor perturbation such as a satellite
companion.

The evolution of the gas velocity field during the massive inflow
period is presented in Figure~\ref{fig5}. The 2-armed feature in the
first panel ($t=0$) clearly illustrates the immediate response of the
gas particles to the presence of similar stellar features.  The gas
particles that are presently unaffected by the arm potential will be
captured by the highly shocked, dense filament within a rotation
period ($\sim 10^7 - 10^8$ years).  Much of the gas particles in the
following snapshot ($t=3.1 \times 10^7$ years) are captured and
trapped in the arm potential, while some of the particles are able to 
stream through the spiral arm which results in 
substantial energy loss.  At $t=6.2 \times
10^7$ years, most of the outer region is void of gas particles except
for the few particles which have streamed through the spiral arm,
or infalling particles from the periphery of the disk.  Finally at
$t=9.4 \times 10^7$ years, an insufficient supply of gas in the outer
disk decreases the rate of the massive inflow, and much of the gas has
already been settled into a ring morphology in the inner kpc.  Despite
the large difference in the orbital geometry, the gas dynamics in the
inner disk appears to be insensitive to the details of the orbital
geometry as evident from the evolution of disk 2 which undergoes a
similar scenario.

\clearpage

\begin{figure*}
\epsscale{0.5}
\caption{Evolution of the rest frame gas velocity vectors from initial 
contact to the end of massive inflow of disk 1 ($left$) and disk 2 ($right$) 
at $t= 0$, $t= 3.1 \times 10^7$, $t= 6.2 \times 10^7$ and 
$t= 9.4 \times 10^7$ years. The vector lengths represent the particle speed.}
\label{fig5}
\end{figure*}

\clearpage

\subsubsection{Semi-Equilibrium Stage: the Formation of a Nuclear Ring ($t \sim 1.0 \times 10^8$ to coalescence)}

While the large scale distribution and kinematics are dominated by the
diffuse gas that continues to funnel toward the centers of the
galaxies, the decreased rate of inflow marks the end of the massive
inflow period in the ring dominated inner disk.  The nuclear gas ring
is commonly observed in central regions of barred galaxies
\citep{buta96} and its origin is often attributed to the presence of a
resonance that introduces a set of orthogonal periodic orbits that
govern the kinematics inside the bar.  The orthogonal orientation of
the ring with respect to the bar resembles orbital structure similar
to the $x_2$ orbits, which indicates the existence of the ILR
\citep{sellwood93}.  However, since detailed orbital analysis is
beyond the scope of the current study, only the morphological evidence
suggest the existence of such resonance. As the disks evolve further,
gas particles in the ring continuously dissipate energy allowing some
of the gas particles to stream further inward.  The resulting
morphology resembles a gas bar near the final stages of the merger
before the two nuclei coalesce.

\subsection{The Properties of the High Density Gas}

\clearpage

\begin{figure*}
\epsscale{1}
\caption{Side by side comparison of the stellar ($left$) and the gas 
($right$) distribution at $t=3.1 \times 10^7$ years.   The length of the 
vertical and horizontal axes is 14kpc, and the same color definitions 
are used as in Figure 2.  The 3.5 kpc radius is shown for reference.  
The labels A and B directly correspond to the same labels in Figure 14.}
\label{fig6}
\end{figure*}

\begin{figure*}
\epsscale{1}
\plottwo{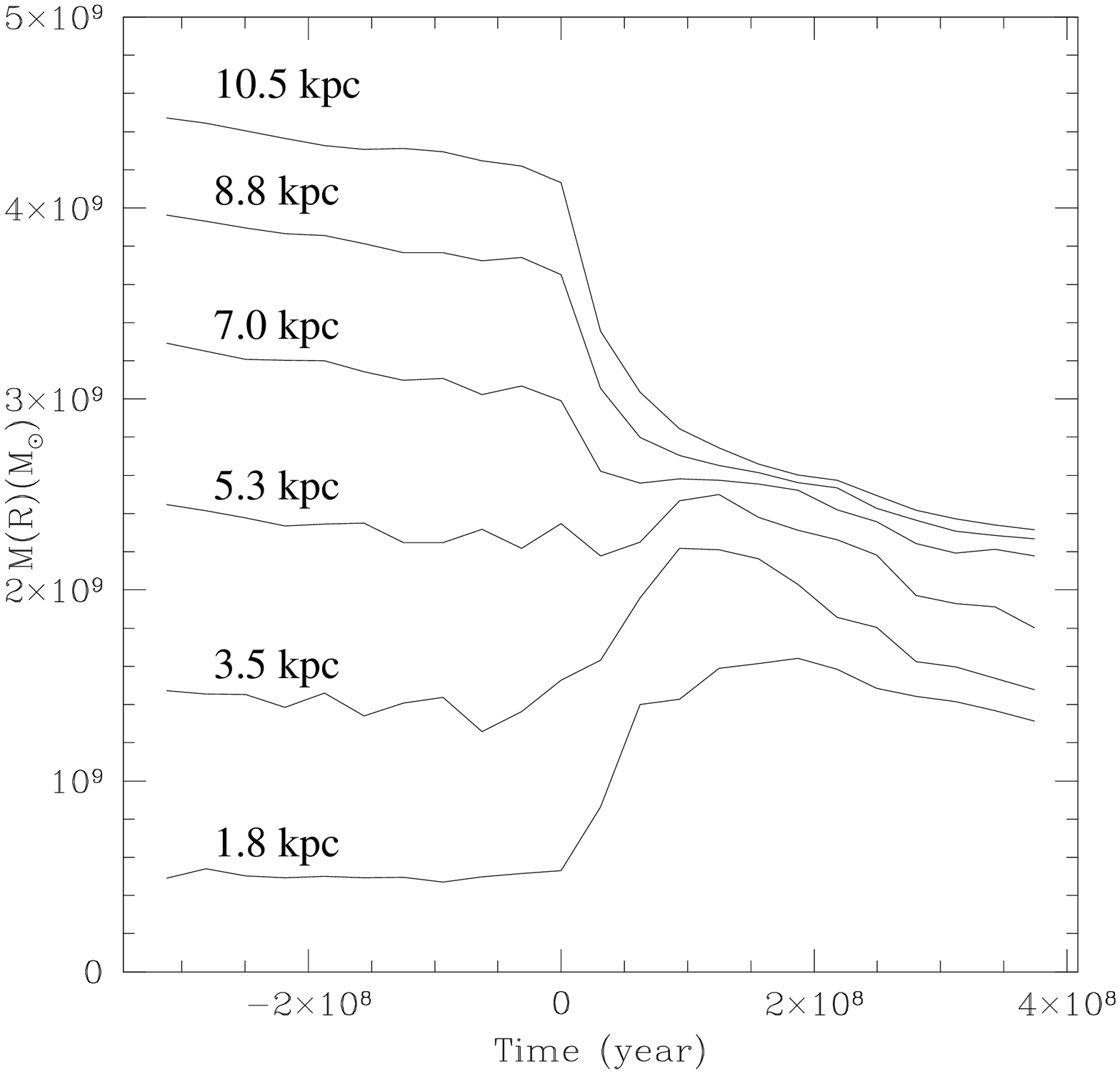}{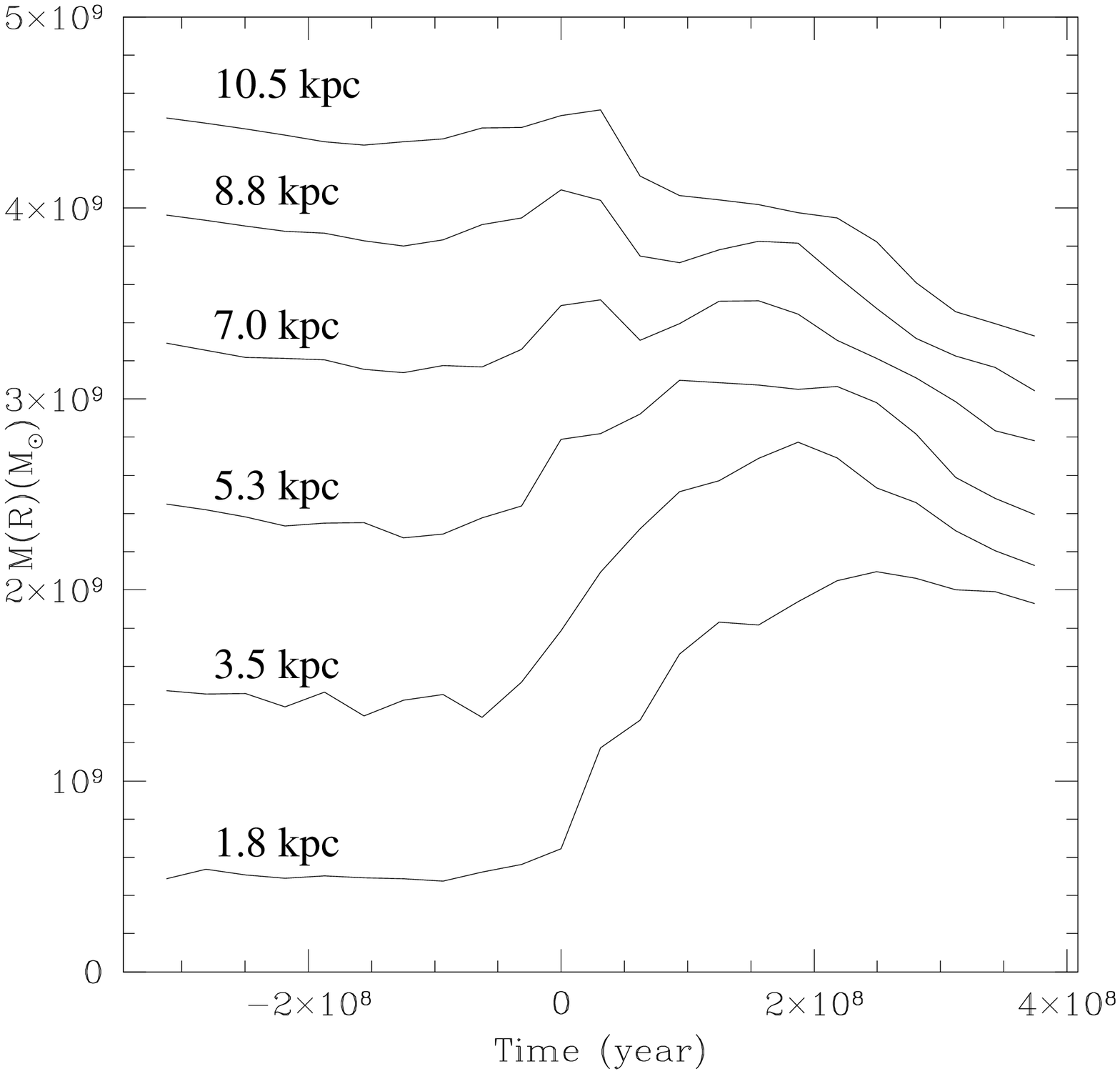}
\caption{Evolution of gas mass within concentric radial bins for disk 1($left$) and disk 2($right$).  Each line represent the cumulative total mass within the labeled radius.}
\label{fig7}
\end{figure*}

\begin{figure}
\epsscale{1}
\plotone{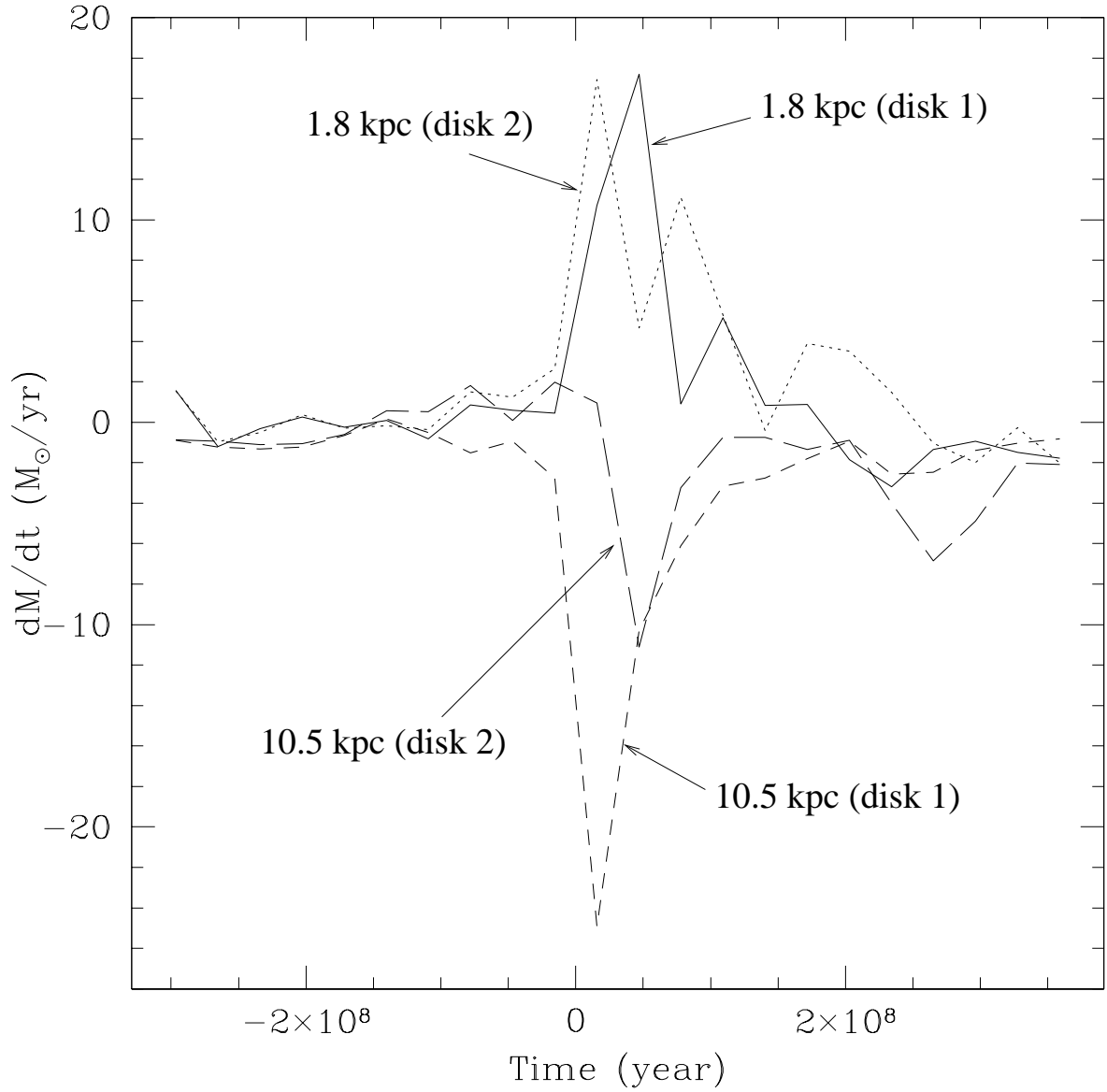}
\caption{The rate of change of gas mass ($dM/dt$) enclosed within a
given radius. Each line represents the gas within 1.8 kpc in disk 1
(thick) and disk 2 (dotted), and within 10.5 kpc in disk 1 (dashed)
and disk 2 (long dashed). Positive values signify inflow across the
boundary, while negative values signify outflow.}
\label{fig8}
\end{figure}

\clearpage

Molecular gas and dust are often used as tracers of shocks.  High
resolution molecular gas observations of nearby galaxies often reveal
a coexistence of dust and molecular gas which occurs most commonly at
the leading edge of the bar \citep[e.g.][]{jogee02}.  In contrast, the
limitation in the particle resolution, dynamic range and the
uncertainties in the artificial viscosity makes it less trivial to
identify shock regions in simulations.  One way to identify shock
region is to search for steep density gradients, however it is a well
known problem that the smoothing nature of SPH complicates the precise
characterization of gas properties when sudden changes in particle
distribution occurs.  An alternative way is to search for
discontinuities in the velocity vectors, which is independent of the
particle distribution as long as the dynamic range is sufficient near
the candidate shock regions.  The arm region in Figure~\ref{fig5}
(labeled A) is a clear example showing velocity discontinuities and
indicating the presence of shocked gas.  During the initial stages of
the inflow, these shocked regions are evident in both the tidal arms
and at the two opposite edges of the forming nuclear ring where the
elongated orbits tend to cross.  The latter may identify the spots
similar to where the bimodal molecular emission is found in isolated
nearby barred galaxies \citep[the molecular ``twin
peaks'',][]{kenney92}.

A side-by-side comparison between the stellar and the gas distribution
shortly after the initial collision ($t = 3.1 \times 10^7$ years) is
shown in Figure~\ref{fig6}.  The stars form a prominent bar with a
major axis of 3.5 kpc and a minor-axis of 1.2 kpc (although we again
caution that the exact extent of the bar will depend sensitively on
the density structure of the hopst galaxy).  While the more diffuse
gas (green; $n_{\rm gas} \geq 1$ cm$^{-3}$) outlines the inflowing gas
and surrounds the nuclear region at a few kpc scale, streams of high
density gas (yellow; $n_{\rm gas} \geq 10$ cm$^{-3}$) are also found,
often corresponding to where discontinuities in the velocity field are
evident.  There is also evidence of very high density gas (red;
$n_{\rm gas} \geq 100$ cm$^{-3}$) at the central region of the galaxy
during the latter periods of the inflow.  The mean densities of the
observed galactic GMCs range from $10^2$ to $10^3$ cm$^{-3}$, however
inspection suggest that most of the simulated gas particles have
volume densities less than $10^2$ cm$^{-3}$ (also see Figure 2).  The
smoothing length is about 50 to 100 pc in the denser nuclear regions,
but the radius of a typical galactic GMC is 20 pc \citep{scoville87}.
Since each SPH particle approximately represents one galactic GMC in
mass, this suggests that the volume filling factor of a typical GMC
enclosed in one SPH smoothing length is only a few percent.  This
correction allows a sensible cut at separating the SPH gas into
different gas tracers according to their smoothed local density.  In
the remainder of this article, we define particles with $n_{\rm gas} =
1$ $cm^{-3}$ or above to represent the gas dominated by dense gas,
while anything below are considered to be in in the diffuse component
of the ISM.  Thus the denser regions (green) may represent sites of
GMCs, and the highest density regions (yellow and red) may represent
GMC complexes and associations.  The latter are possible sites of
current star formation, the intensity of which depends on the degree
of shear experienced by the high density gas.  On the other hand, the
diffuse gas (blue) may identify the tidal tails traced observationally
by atomic hydrogen emission.

\subsection{The Inflow Rate}

Figure~\ref{fig7} presents the differential evolution of the
cumulative gas mass enclosed within concentric rings around the disk
center. Radial and azimuthal gas compression begin to convert the
higher density gas into stars, decreasing the overall gas mass slowly
before the collision, while violent inflow and ejection into tidal
tails rapidly changes the overall distribution of gas after
pericentric passage. It is clearly seen that the stronger response in
the $outer$ disk is more evident in the prograde disk (disk 1), while
relatively smaller mass loss is seen in the high inclination disk
(disk 2).  At the final stages of the massive inflow ($t \sim 10^8$
years), the gas in disk 2 is $40\%$ more abundant than in disk 1
inside a radius of 10.5 kpc ($80\%$ of gas is initially distributed
within this radius).  The prograde geometry of disk 1 forces ejection
of much of the gas ($1.0 \times 10^9$M$_\odot$) out to the tidal tails
and the bridge medium.  Shortly before the coalescence, about $30\%$
$(2.7 \times 10^8$M$_\odot$) of the gas has returned to disk 1, but
$60\%$ $(5.9 \times 10^8$M$_\odot$) of the ejected mass from disk 1
has been captured by disk 2, $30\%$ $(2.0 \times 10^8$M$_\odot$) of
which is captured almost immediately ($< 10^8$ years) after the
collision.  The remaining $10\%$ hovers around the bridge region until
the final coalescence.

While the mass evolution in the outer disk strongly depends on the
orbital geometry due to the presence of the spin-orbit resonance
operating on the loosely bound particles, the evolution of the {\it
inner} disk is less affected by the resonance and is dominated more by
the density structure of the progenitor galaxy.  Overall, both disks
show a central inflow ($R \leq 1.8$ kpc) with a total gas mass of a
$(1-2) \times 10^9$M$_\odot$ before the galaxies coalesce, and this
behavior is independent of the orbital geometry.  Figure~\ref{fig8}
presents the evolution of the inflow/ejection rate at the inner (1.8
kpc) and outer (10.5 kpc) most radii in Figure~\ref{fig7} shown here
for both disks.  The mass inflow rate at R = 1.8 kpc shows similar
evolution in both disks because the evolution of the inner disk is
less sensitive to the orbital geometry, while the outer disk is more
susceptible to the resonance affecting the prograde disk (disk 1) more
severely. At R = 10.5 kpc, the average ejection rate is 9 and 3
M$_\odot$/yr with a maximum found just after the initial collision at
25 and 11 M$_\odot$/yr for disk 1 and disk 2 respectively.  At R = 1.8
kpc, the average inflow rate is 7 and 8 M$_\odot$/yr for disk 1 and
disk 2, with both disks showing a maximum value of 17 M$_\odot$/yr.
These inflow rates are comparable to those derived from observations
of isolated and interacting barred galaxies \citep[$0.1-10$
M$_\odot$/yr;][]{jogee99,sakamoto99b,petitpas02}, although significant
uncertainties exist when trying to apply the simulation results to
individual observed galaxies.  Nonetheless, the large inflow rate
calculated here is more than enough to initiate and sustain central
starburst activity at a few M$_\odot$/yr, but may not be high enough
to account for the very high star formation rates observed in ULIRGs
($\geq 100$ M$_\odot$/yr) \citep{sanders96}.

\subsection{The Bulgeless Case}

MH96 demonstrated that a collision of bulgeless galaxies results in a
strong stellar bar that dominates more than half of the radial extent
of the entire galaxy, and produces high inflow and star formation
rates shortly after the initial encounter.  While galaxies with a
central bulge dominate the luminous galaxy population in the nearby
universe at least up to $z \sim 1$ \citep{lilly98}, increasing number
of bulgeless galaxies are suggested in the early universe
\citep{wyse97}.  Thus multiple collisions between bulgeless galaxies
may be a more common phenomenon in the early universe, and may also
play an important role in the early stages of galaxy formation and
evolution.  Figure~\ref{fig9} presents the density evolution of the
gas disk in the bulgeless galaxy simulation similar to that in
Figure~\ref{fig2}, but shown here up to $t = 2.6 \times 10^8$ years
because the inflow timescale is longer by a factor of 2.  The absence
of the bulge permits the growth of the stellar bar to a size of about
10 kpc, which is 3 times longer than what is formed in the bulge
dominated galaxies (see the comparison in Figure~\ref{fig10}).  High
density gas ($n_{\rm gas} \leq 10$) is completely absent initially ($t
< 0$) in the central region.  Only after $\sim 5 \times 10^7$ years
after the collision does azimuthal compression of gas raise the
smoothed local densities of the gas to a higher level ($n_{\rm gas}
\leq 10$) in the densest regions of the bar-dominated galactic disk.
The subsequent evolution of the gas is also clearly different from the
bulge dominated case, forming a linear bar-like structure that allows
direct central streaming along the strong stellar bar. A long strip
(length $\sim 6$ kpc) of very high density gas ($n_{\rm gas} \geq
100$) forms as a consequence of the gas accumulation toward the
central region.  Finally, the inflow of dense gas effectively declines
at $t \sim 2.0 \times 10^8$ years.  The inflow rate to the inner 1.8
kpc in the initial stages of the inflow period is much lower than the
bulge dominated case ($\sim 5 M_{\odot}$/yr for both disks), but peaks
at $1.6 - 2.1 \times 10^8$ years at 25 and 14 $M_{\odot}$/yr for disk
1 and disk 2 respectively.

This result clearly demonstrates the sensitivity of the behavior of
the gas to the detailed stellar structure that could profoundly change
the resulting evolution of the gas and ultimately to the distribution
and the intensity of the subsequent starburst activity.  Some clues as
to why the behavior of the gas differers dramatically between the two
models may be obtained by comparing the results with the
hydrodynamical simulations in a steady-state bar potential
\citep[e.g.][]{athanassoula92a}, and it appears to be intimately
linked to the properties of the ILR.  These models suggest that the
physical properties of the bar dictates the presence/absence of the
ILR; i.e. the ILR is absent when the bar has large axial ratio or low
central concentration.  The resulting stellar properties found in the
bulgeless models fulfill these conditions at least qualitatively, but
a thorough quantitative description is beyond the scope of the present
paper and will not be conducted.  Nevertheless, if the structural
properties of the bar indeed govern the existence of the ILR and the
orbital structure of the gas, the existence of the ILR (and
consequently $x_2$ orbits) in the bulge dominated galaxies effectively
prevents the gas from further reaching the very center of the galaxy,
as the gas particles streaming along the nuclear ring suffers from
little torque and dissipation.  This may explain the less intense star
formation rate seen in the bulge dominated disk encounter model at the
stages following the initial collision (MH96).  On the other hand, for
a galaxy with no ILR, gas streams directly to the central region
without interference.  The sharp rise in star formation activity in
the bulgeless model may be attributed to such direct central inflow.

In summary, a few markedly distinct properties are evident when a
comparison between the two cases are made.  First and most
importantly, the inflow in the bulgeless galaxy is characterized by a
direct central streaming where a linear bar-like structure dominates
the distribution and the kinematics of the inflow. In contrast, the
bulge dominated galaxy forms a nuclear ring that largely delays the
central inflow and the subsequent star formation activity.  Secondly,
 the gas in bulgeless galaxy 
seems to take slightly more time to respond, likely due to the 
different internal structure and dynamical timescales.
The initial inflow rate is much smaller in the bulgeless galaxies which
therefore results in a longer inflow timescale ($t \sim 2.0 \times
10^8$ years) compared to the bulge dominated disk encounter ($t \sim 1.0
\times 10^8$ years).  The peak inflow rates are comparable at 
(15-25)~$M_{\odot}/yr$.  By comparing the gas morphology to the results
from hydrodynamical simulations of \citet{athanassoula92a}, it is
suggested that these two are both possibly related to the
characteristics of the stellar bar, whose properties can profoundly
alter the governing orbital structure of gas within the bar potential.

The linear appearance of the gas in the bulgeless galaxy is remarkably
similar to that of the so called ``chain galaxies'' observed at high
redshifts \citep{cowie95, taniguchi01, elmegreen04}.  These are a
population of galaxies that have axial ratio of $\sim 5$
\citep{taniguchi01}, and has been speculated to be star forming
galaxies at intermediate to high redshift ($z\sim 0.5 - 3$)
\citep{cowie95}.  The observational evidence of increasing number of
bulgeless galaxies in the early universe, and the linear morphology of
gas inflow predicted in this study offers a possible new
interpretation to the origin of the ``chain galaxy'' phenomenon;
forced radial inflow and linear star formation in bulgeless galaxies.
It is possible that this proposed scenario may not necessarily require
a presence of an equal mass perturber as massive bulgeless disks are
unstable to bar formation intrinsically \citep{ostriker73, oneill03},
but the intrinsic perturbation will be significantly smaller and the
amplitude of the resulting bar may not be as extreme.  This transient
phase persists for $\sim 10^8$ years which is much longer than the
timescale for the massive UV-emitting young stars to form ($\sim
10^{5-6}$ years), assuming that the gravitational collapse can
withstand the immense radial shear from the inflow.  In these systems,
the kinematics of the gas is dominated by the inflow and any attempts
to derive the rotation velocity will substantially underestimate the
true rotation of the galaxy even when the galaxy is viewed edge-on.
This is consistent with the finding by \citet{bunker00} that HDF
4-555.1 (a ``chain galaxy'' at z = 2.80) has a small rotation velocity
($V_{rot} < 100$ km/s), and systematic spectroscopic surveys of the
``chain galaxy'' population should reveal a better picture of its
origin in the future.

\section{The Structural Parameters: the Asymmetry, Concentration and Compactness of Stars and Gas}

The development of sensitive detectors mounted on space-borne
observatories allow deep and large scale galaxy surveys whose sample
size now exceeds the limit in which one can investigate each source
individually.  Because it is a more systematic and robust way to
measure galaxy morphology, automatic classification of the optical
galaxy morphology has become increasingly popular in recent years.
Two structural parameters that are commonly computed are the asymmetry
parameter ($A$) \citep{schade95, abraham96, conselice00, bershady00},
and the concentration index ($C$) \citep{morgan58, fraser72, doi93,
abraham94, graham01, conselice03}.  The application of the structural
parameters to gas morphology has not been done in the literature to
date primarily because of the lack of spatial resolution that has only
become readily available recently with the use of interferometric
techniques.  In addition, morphological classification of ISM in
galaxies has not received much attention in the context of
understanding galaxy formation and evolution.  The observational
sample presented in paper II, as well as recent CO~(1--0) surveys of
nearby galaxies (i.e. BIMA SONG \citep{helfer03}) are obtained at high
enough spatial resolution ($\leq 1$ kpc) and S/N making such
investigations feasible.  In what follows, similar structural
parameters, $A$ and $C$ are computed for the simulated morphology of
both the stars and gas in order to form a theoretical template that
will aid to understand the structural parameters derived for the
observational sample in paper II.  In addition, in order to better
trace the relative sizes of the gaseous and stellar components of
galaxies, a new parameter $K$, the compactness parameter, is
introduced.

\subsection{Definition of A, C, and K Parameters}

The asymmetry parameter ($A$) is defined as;
\begin{equation}
A = \frac{|I - I_{rot}|}{2I_{tot}},
\end{equation}
where, $I$ is the face-on image of stars (or gas), $I_{rot}$ is $I$
rotated $180^\circ$ around the center of the galaxy (typically defined
as the maxima of stellar density), and $I_{tot}$ is the total
integrated intensity of a single disk.  Uncertainties in the centering
is the dominant source of error in the asymmetry parameter, and the
common approach that has been adopted in observation is to use the
global minima of $A$ after iterating through each pixel in the image.
This method will likely result in a rotation center that is close to
the dynamical center of the disk as long as the stellar or gas
distribution is azimuthally and radially near uniform.  However, the
true interpretation of $A$ becomes much more complex when two or more
galaxies are involved in the determination of a single $A$, and only
the evolution of $A$ for a single galaxy in the pair is computed here
to avoid confusion.  To best represent the observational procedure in
paper II in which the rotation center is defined from the peak
coordinates in the relatively extinction-free K-band images, the disk
potential minima is used here to define the rotation center of the
galaxy.  The above definition normalizes $A$ so that $A = 0$ for
complete axisymmetry and $A=1$ for total antisymmetry.  The asymmetry
parameter is computed for both the raw image whose pixel resolution is
216 pc/pixel, and the image convolved with a Gaussian beam (FWHM = 1
kpc) allowing a direct comparison to the observational sample in paper
II which has a mean spatial resolution roughly 1 kpc in FWHM of the
synthesized beam.  This and all of the stellar structural parameters
derived in the remainder of this article assumes that the
mass-to-light ratio does not change across the disk.

The concentration index ($C$) is defined as the ratio between the outer 
and inner regions represented by the radius enclosing 20 and 80\% of the 
total flux, or;
\begin{equation}
C = \frac{r_{80}}{r_{20}},
\end{equation}
where $r_{20}$ represents the radius (i.e. $r_{20} = \sqrt{NA_p/\pi}$ 
where $N$ = number of pixels, $A_p$ = area of each pixel) enclosing 20\% 
of the total emission from the peak, and $r_{80}$ is the same but enclosing 
80\% of the total emission.  The concentration index is calculated for 
both stars and gas for comparison.

The new compactness index ($K$) quantifies the relative concentration of 
gas with respect to the size of the stellar disk;
\begin{equation}
K = \frac{r^s_X}{r^g_X},
\end{equation}
where $r^s_X$ and $r^g_X$ represent the radius enclosing X\% of the total 
flux of stars and gas respectively.  The stellar radius ($r^s_X$) is  
sensitive to the the value adopted for $X$; $r^s_X$ shows small evolution 
for small X since the particles inside the tidal radius are less affected 
by the companion tidal force, while particle ejection into the tidal tails 
increases $r^s_X$ rapidly for large X (Figure~\ref{fig11}).  On the other 
hand, radial streaming and central condensation of the gas particles makes 
the gaseous radius ($r^g_X$) less sensitive to the value adopted for $X$.

The evolution of the A, C and K parameters for gas are calculated
using two different density criteria; (1) \textit{all} gas particles
confined within a 14 kpc radius to investigate the behavior of both
the diffuse and dense gas, and (2) all gas particles above $n_{\rm
gas} = 1.0$ (see Figure 2; green) to trace only the denser component
of the gas.  Creating stellar maps with different density thresholds
is more difficult to implement in a meaningful way, since it requires
the use of a model for the mass-to-light ratio and particle smoothing,
both of which are significantly nontrivial. As such, we only measure
the quantitative morphological parameters for the stellar distribution
as a whole.

\subsection{The Evolution of A, C and K Parameters}

The evolution of A, C and K for both stars and gas are presented in
Figures~\ref{fig12} and \ref{fig13}. The stellar asymmetry ($A$)
increases by a factor of two over a period of $\sim 7 \times 10^7$
years and maintains approximately a similar amplitude until the end of
the sequence presented here.  The spatial smoothing results in only a
modest reduction in $A$ since the local and central condensations seen
in the gas are absent in collisionless stellar particles.  The small
dynamic range in $A$ simply reflects the intrinsic axisymmetric nature
of the tidally induced features, and blindly applying this technique
to examine the degree of tidal disturbance therefore can result in
large uncertainties.  In reality, the angular resolution and line of
sight projection can largely complicate the application of this
technique to a single galaxy in a pair, and a more practical way is to
derive a single value of $A$ to the system as a whole, similar to what
has been demonstrated in high resolution optical images of galaxies
\citep[e.g.][]{conselice03}.  The resulting $A$ in an intermediate
stage merger in this case may be sensitive to the relative stellar
mass ratio of the pair assuming a constant M/L ratio, since the
asymmetry is sensitive to the residual structure in the most luminous
parts of the galaxies (i.e. the bulge).  These results suggest the use
of $A$ as a replacement for the Hubble classification scheme is
somewhat technically ambiguous even when a single galaxy in a pair is
tracked instead of the system as a whole.  The situation only becomes
worse at higher redshift when surface brightness dimming, loss of
angular resolution and line of sight projection effects all play a
role in further complicating the interpretation of the analysis.
 
The modest dynamic range seen in the evolution of the $A$ parameter
for the dense gas particles further supports the idea that the use of
$A$ to trace the dynamical evolution is mostly impractical.  A $\sim
50\%$ growth in $A$ after $(5-10) \times 10^7$ years is seen but the
increase is too small to be effectively used as part of a distinct
classification scheme.  In contrast to stars, the Gaussian smoothing
of the gas significantly reduces the clumpiness in the distribution,
reducing $A$ by about a factor of two overall.  The initial value of
$A$ when all of the gas particles are used is comparable to the case
in which the dense gas is smoothed to a 1~kpc beam, but the evolution
subsequent to the collision becomes similar to that of the unsmoothed,
dense gas particles.  The gas and stars both initially follow a smooth
exponential radial decline, which leads to the slightly evolved
progenitor disks with near azimuthal symmetry.  Such an idealized
scenario will inevitably result in a gas response with a strong axial
symmetry since the underlying stellar disk responds to the collision
with a strong $m=2$ perturbation, but weak in $m=1$ (\S3.1.2).  A more
realistic case will involve collisions of disks where the distribution
of gas is initially more clumpy and filamentary, in which case the
excitation of the stellar $m=2$ mode may result in a gas response that
largely depends on the initial distribution, suggesting that the
results presented here are only lower limits.  In concordance with
such speculation, the initial investigation of $A$ applied to the
isolated galaxies in the BIMA SONG sample \citep{helfer03} yield a
mean of $0.47 \pm 0.26$ (unsmoothed) and $0.29 \pm 0.16$ (smoothed to
the worst resolution in the sample), and $0.50 \pm 0.22$ and $0.35 \pm
0.21$ for the colliding galaxy sample in Paper II respectively.

The evolution of $C$ is sensitive to the behavior of $r^s_{80}$ and
$r^g_{80}$ since the inner region of both stars and gas ($r^s_{20}$
and $r^g_{20}$) responds to the collision with significantly smaller
amplitudes.  Thus the ejection of the outer stars into tidal tails is
directly seen as a monotonic increase of C in stars.  Gas particles on
the other hand exhibits a small peak shortly after the collision that
is followed by an immediate drop which is simply a reflection of the
behavior of gas at $r^g_{80}$.  This behavior is seen in the case
where all of the gas particles (i.e., not just dense gas) is used, but
to a smaller degree.  These results suggest that $C$ is a poor tracer
of structural evolution of gas during a major merger, but its
application to stars alone appears to work reasonably well.

The strong sensitivity of $K$ to the amount of flux included (i.e. the
value of $X$) illustrated the differences in how various parts of the
disk evolve.  The stellar disk expands rapidly and monotonically
outside the tidal radius (between $X=50$ and $68$), but the evolution
of the dense gas is more transient, marked by an initial increase
followed by a rapid decline in both $X=50$ and 68 (see
Figure~\ref{fig11}).  The ratio of these ($X=68$), therefore, results
in a slow initial evolution followed by a stronger evolution toward
the latter epochs of the massive inflow period.  A similar trend is
seen in the case where all of the gas particles are traced, but the
post-collision value of $K$ is smaller by a factor of two.  One
potential problem concerning the use of $K$ is when the observation is
limited by angular resolution and sensitivity.  A simple test using a
Gaussian shows that the galaxy needs to be extended at least 8 pixels
in FWHM, otherwise the ratio is overestimated by a factor of a few.
Another potential problem arises when the data are not sensitive
enough to detect the fainter tidal structure in stars.  Thus the
idealized results presented here may only provide an upper limit to
$K$ since $r^s_{68}$ may be significantly smaller when the extended
emission is not accounted for.  Since further discussion depends on
the quality of the actual data, these issues are further addressed in
detail in the analysis of observational data in paper II.

\clearpage

\begin{figure*}
\epsscale{1}
\caption{Similar to Figure 2 but showing disk 1 of a $bulgeless$ encounter.}
\label{fig9}
\end{figure*}

\begin{figure*}
\epsscale{1}
\plottwo{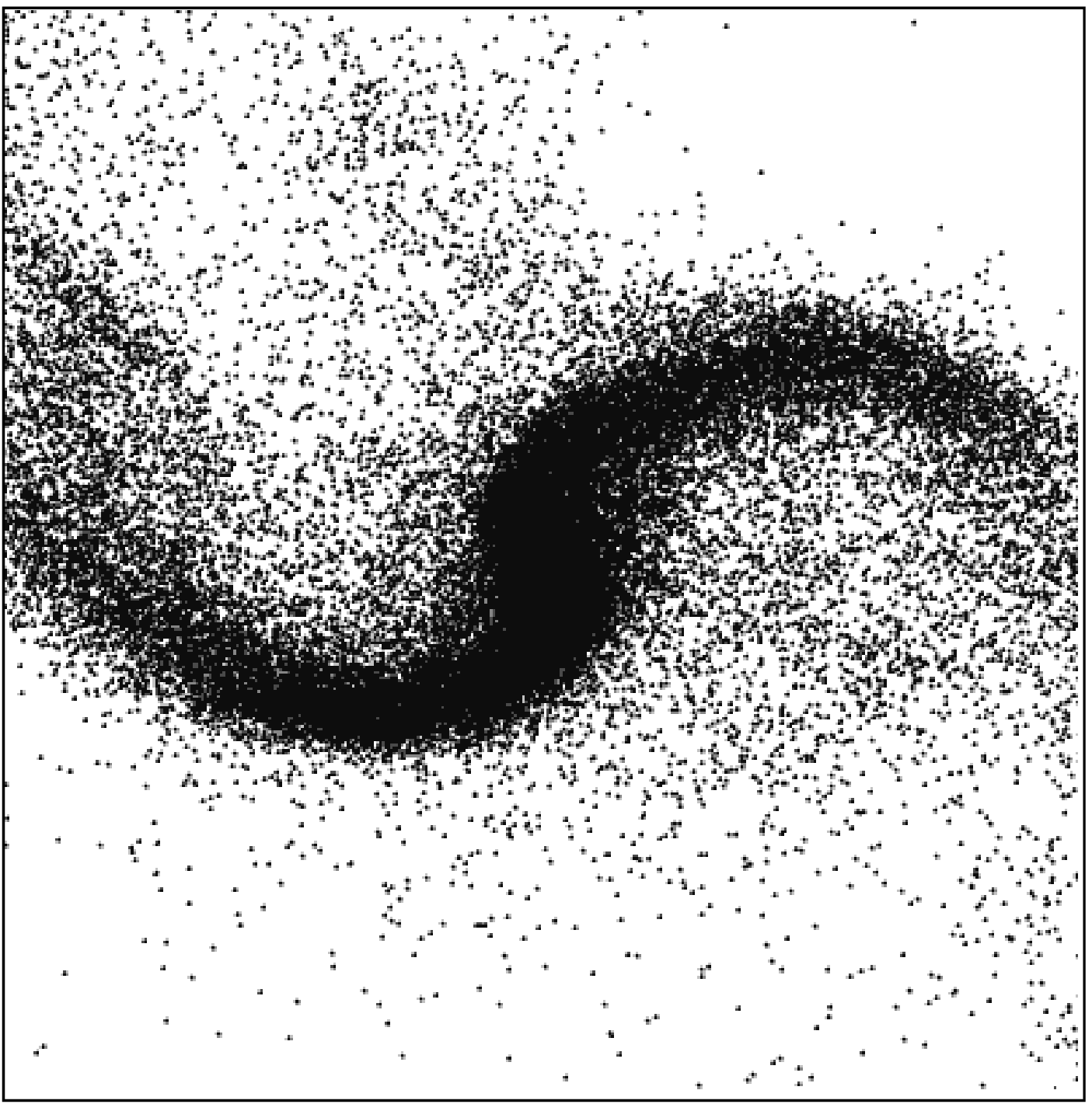}{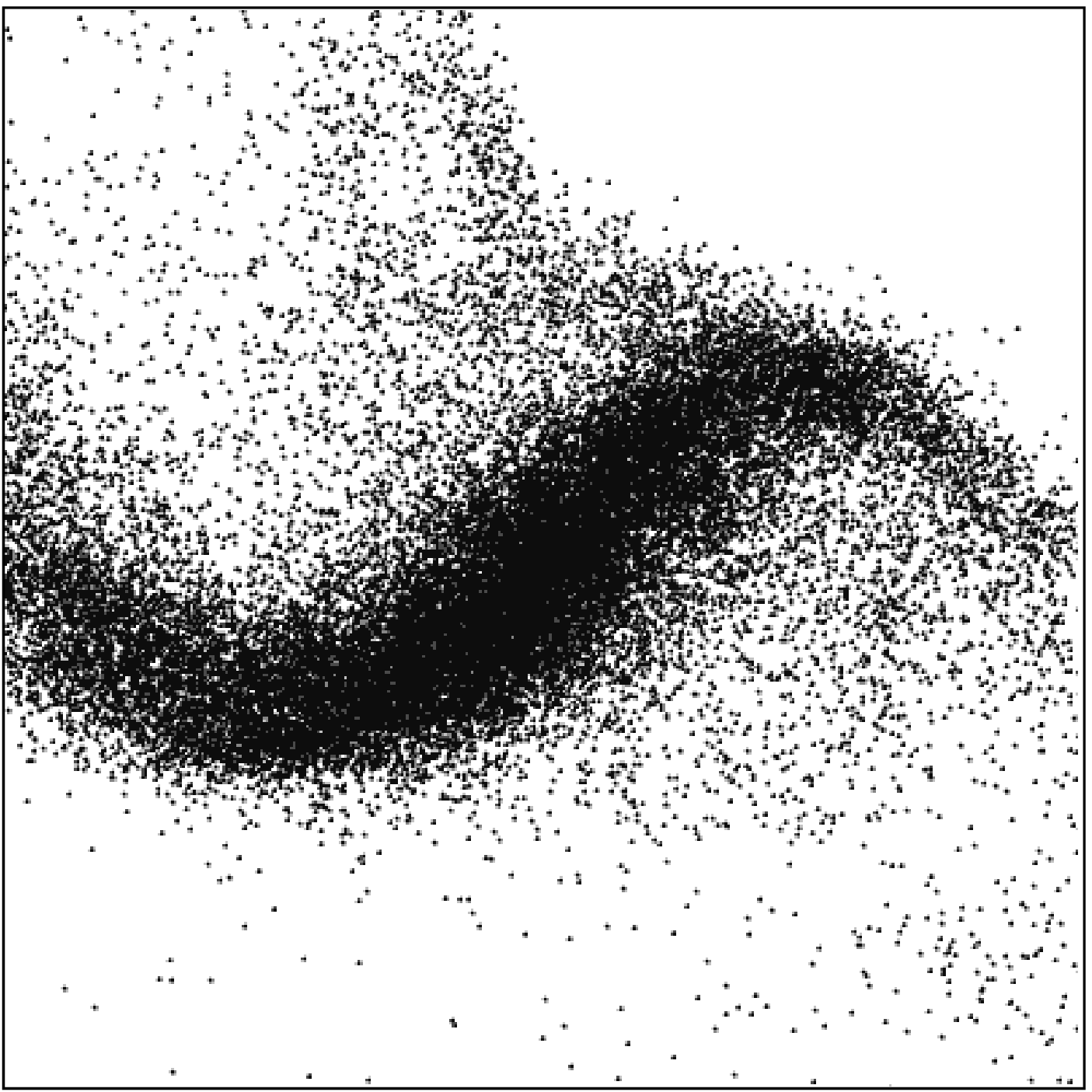}
\caption{Comparison of the stellar distribution between the ($left$)
bulge dominated galaxy at $3.1 \times 10^7$ years and ($right$)
bulgeless galaxy at $5.2 \times 10^7$ years.  The boxes are 28 kpc on
a side.}
\label{fig10}
\end{figure*}

\begin{figure*}
\epsscale{1}
\plottwo{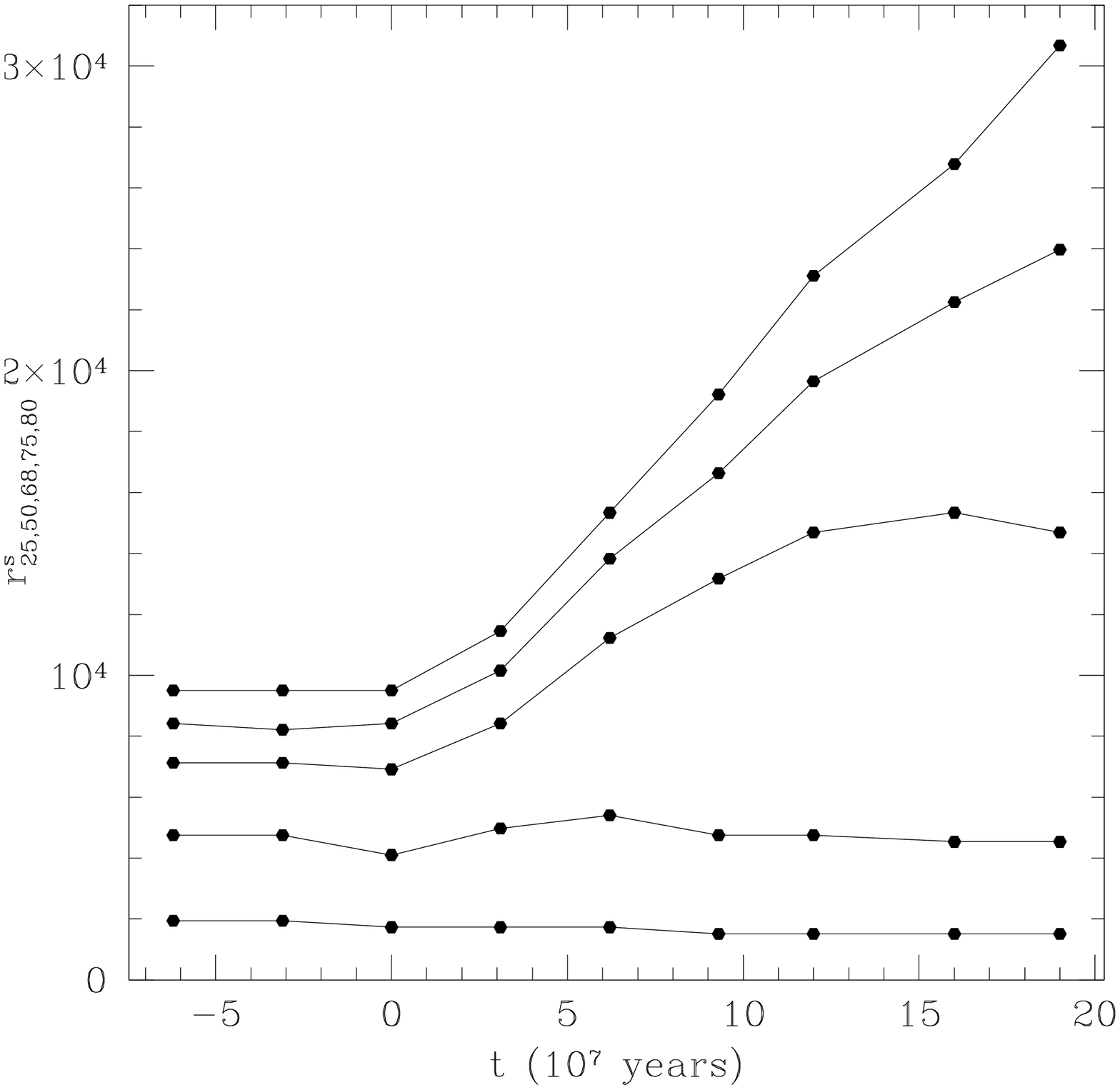}{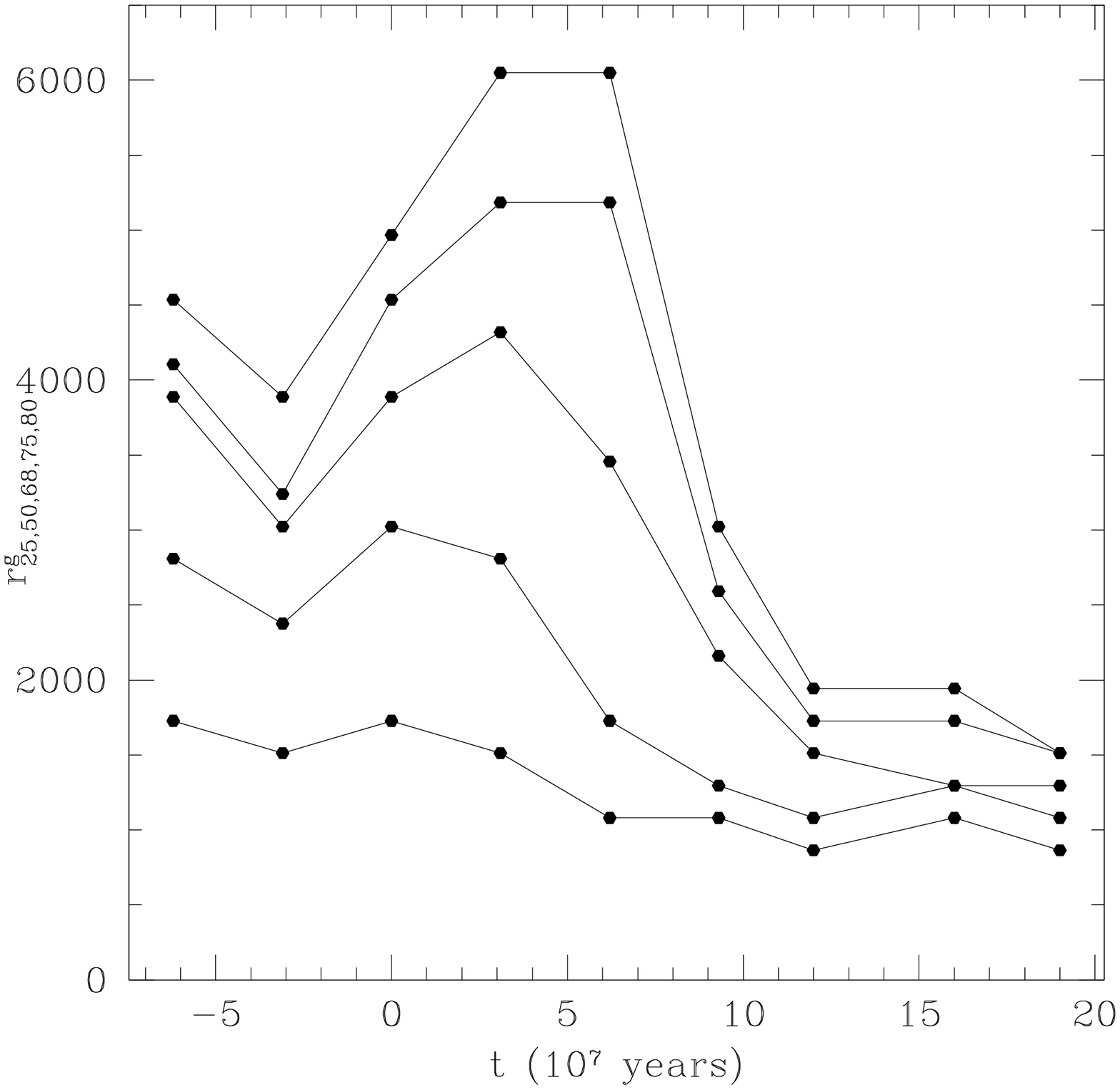}
\caption{(\textit{left}) Time evolution of the stellar extent $r^s_X$,
where $r^s_{25}$,$r^s_{50}$,$r^s_{68}$,$r^s_{75}$, and $r^s_{80}$ are
plotted from bottom to top.  The units are in parsecs.
(\textit{right}) Same as in \textit{left} but for the gas particles.}
\label{fig11}
\end{figure*}

\begin{figure*}
\epsscale{1}
\plottwo{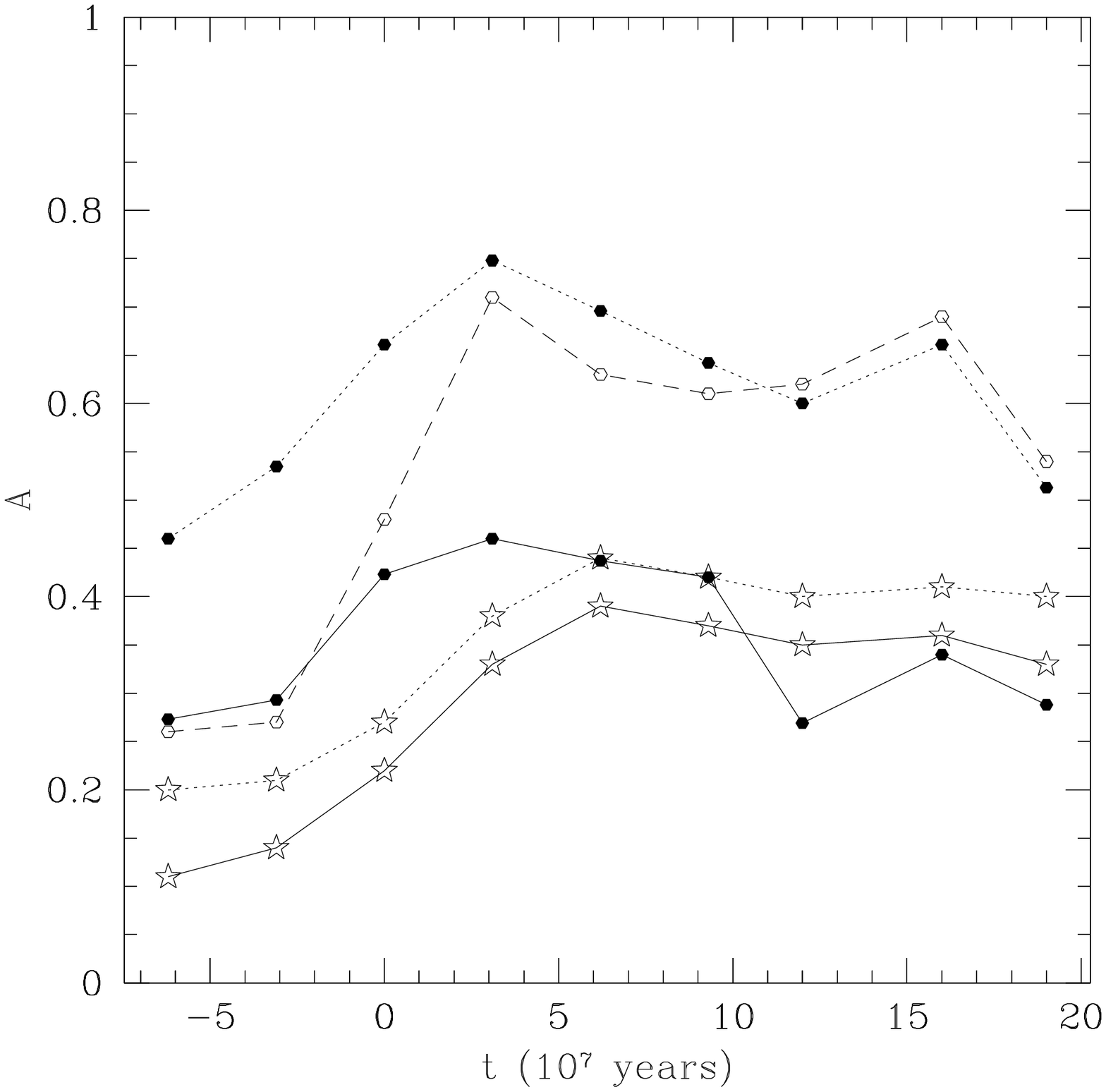}{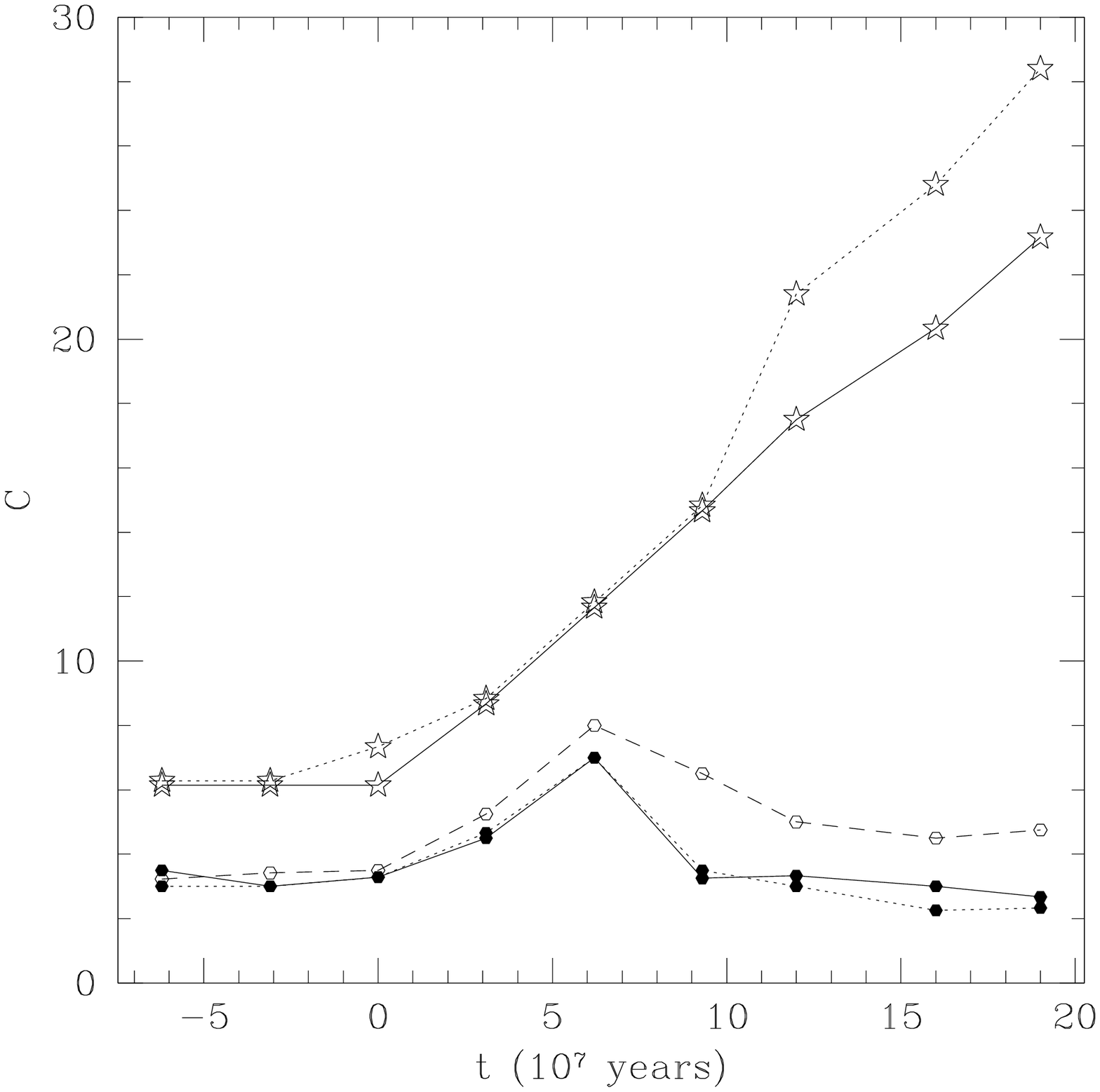}
\caption{($left$) The time evolution of the asymmetry parameter ($A$) 
for the dense gas ($n \geq 1.0 cm^{-3}$) (dotted line with filled circles) 
and the same gas but convolved with a Gaussian with FWHM = 1 kpc (solid 
line with filled circles), and stars (dotted lines with star symbols) and 
stars convolved with a Gaussian with FWHM = 1 kpc (solid line with star 
symbols).  Also shown is the evolution of $A$ using all of the gas 
particles in the disk (dashed line with open circles).  $A$ is normalized 
such that the value 1 (0) corresponds to complete antisymmetry (axisymmetry). 
(\textit{right}) The time evolution of the concentration index ($C$).  
Large $C$ represents a larger concentration of that specie.}
\label{fig12}
\end{figure*}

\begin{figure}
\epsscale{1}
\plotone{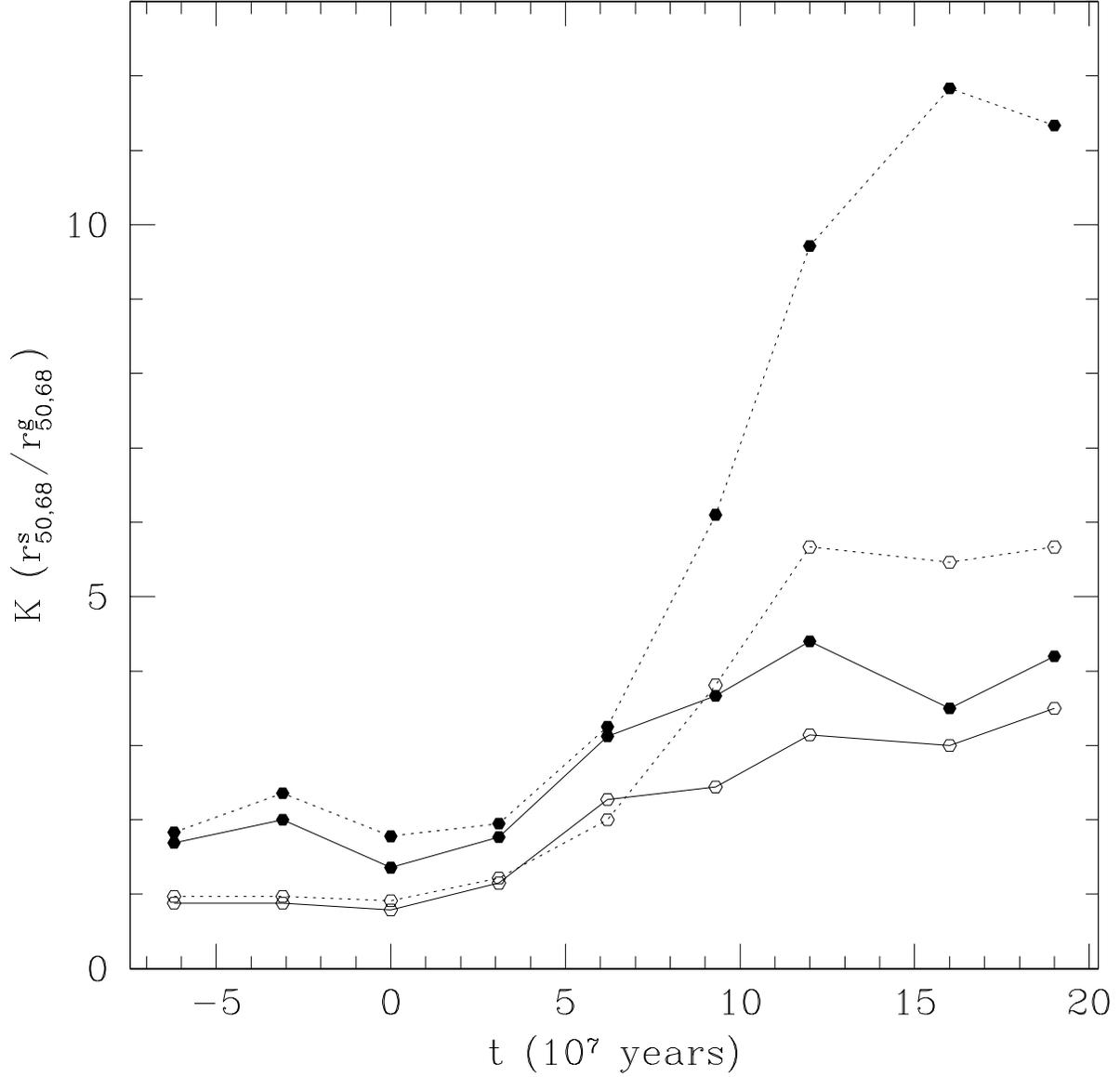}
\caption{Time evolution of the compactness parameter ($K$) calculated
for the dense gas ($n \geq 1.0 cm^{-3}$), where the solid and dotted
lines with filled circles represent $K(r^s_{50}/r^g_{50})$ and
$K(r^s_{68}/r^g_{68})$ respectively. The open circles represent the
evolution of the $K$ parameter using all of the gas and star
particles.}
\label{fig13}
\end{figure}

\clearpage

\section{Implication on the Analysis of Observational Data}

The position velocity diagram (PVD) offers an efficient and powerful 
way for studying the disk kinematics in observations of  molecular and 
atomic gas \citep[e.g.][]{sofue95}.  One of the potential application 
of the PVD is to infer the presence of a  bar in edge-on isolated galaxies 
by identifying anomalous non-circular motions driven by the barred potential 
\citep{athanassoula99,bureau99}.  Therefore, the evolution and the shape 
of the PVD contains important diagnostic information regarding the inflow 
kinematics and this can further help the observational interpretation of 
interacting systems.

\subsection{The Evolution of the PVD}

The evolution of the PVD from $t = 3 - 12 \times 10^7$ years is
presented in Figure~\ref{fig14}.  The PVD slit is aligned with the
major axis of the edge-on disk, viewed from angle 1 in
Figure~\ref{fig2}.  The center of the PVD is fixed at the disk
potential minima, and the same spatial (220~pc) and velocity
resolution (26~km/s) were chosen as before.  In addition, in order to
further investigate how the viewing angle affects the shape of the PVD
in the initial stages of the massive inflow, an additional viewing
angle from $45^\circ$ south (from viewing angle 2 in
Figure~\ref{fig2}) is adopted (Figure~\ref{fig15}).  In both viewing
angles, different cutoff densities are used to study the PVD that is
dominated by different component of the gas.  For example, the PVD
generated using \textit{all} of the gas particles may represent both
the diffuse and dense components of the ISM, while a cutoff density of
1.0 and 10 may filter out the diffuse component leaving only the
regions dominated by the denser clouds.

The detailed structure seen in the PVD is highly dependent on the
viewing angle toward the bar in the initial stages of the inflow
(compare first and second row of $n_{\rm gas} \geq 1.0$ in
Figure~\ref{fig14}); the inflow gas is characterized by distinct
clumps at $t = 3.1 \times 10^7$ years (labeled A and B in
Figure~\ref{fig6} and Figure~\ref{fig14}), while a long and dense
strip in the PVD depicts the inflow gas at $t = 6.2 \times 10^7$ years
(labeled C and D in Figure~\ref{fig14}).  Toward the final stages of
the inflow and near the semi-equilibrium stage, the emission
morphology becomes less sensitive to the adopted viewing angle since
the disk is composed of only two major components (i.e. a dense
nuclear ring and diffuse extended tails). The diffuse gas in the
outskirts of the galaxies (see the left column of Figure~\ref{fig14})
shows a decline in the radial velocity along the line of sight, which
is often seen observationally in extended \ion{H}{1} disks, and is
attributable to warping of the outer disk.

The dense gas forms a morphology that resembles an $\int$-shape
(integral-shape) in all PVDs in Figure~\ref{fig14}, where the
upper-right/lower-left quadrant represents the redshifted/blueshifted
gas along the line of sight.  Careful observation of the inner PVD reveals
that the $\int$-shape begins relatively flat (i.e. the flat part of the 
rotation begins  $r \sim 1.5$ kpc at t=$3.1 \times 10^7$
years) but becomes steeper in the subsequent epochs (i.e. the flat
part of the rotation rotation begins $r \sim 0.8$ kpc at t=$6.2 - 9.3
\times 10^7$ years) and flattens again at the final epoch of the
sequence.  This apparent steepening of the inner rotation is due to
the changing line of sight as the galaxy rotates, and occurs when the
line of sight intersects the stellar bar near end-on, where the
particular orientation projects the gas particles moving
\textit{closer} to the nucleus and consequently \textit{faster}.  On
the other hand, when the bar is observed near edge-on, the projected
distribution of the gas with higher column density appears near the
two edges of the stellar bar where the gas is moving \textit{farther}
from the nucleus and \textit{slower}, resulting in a flatter
$\int$-shape.  This demonstrates one way the emission in the PVD can
be used to predict the approximate line-of-sight orientation of the
bar in a disk observed near edge-on \citep{binney91}, which is
especially useful when the line of sight inclination and extinction
prohibits direct identification of the bar in the optical and NIR
images.

\subsection{Kinematic Signatures of Gas Inflow Found in the PVD}

The large scale PVD morphology of the dense gas (n$_{\rm gas} \geq
10$) generally shows axial symmetry across the origin, while the small
scale details (i.e. darker gray scale in Figure~\ref{fig14}) often
show larger asymmetries.  The example shown in Figure~\ref{fig14} has
stronger emission in the lower left quadrant along the rising part of
the rotation.  This indicates that the dense material is slightly
lopsided toward one side of the galaxy, and that there exists a
depression of nuclear gas (i.e. a central cavity) \citep{sakamoto99}.
Since the central cavity is a sign of the nuclear ring --- which is
shown in \S3 to be a possible consequence of a central inflow ---
observational evidence of such features in the PVD is a potential
signature of past or ongoing inflow.  For this method to work
effectively, the observational images require enough angular
resolution that the nuclear structure is resolved.  Figure~\ref{fig16}
presents the PVD of UGC~12915 (the ``Taffy Galaxy'')\citep{cond93},
demonstrating one such example where a lopsided PVD emission and the
emission in the forbidden velocity quadrant (see below) both suggest
that some of the gas is currently involved in an inflow.

The comparison between Figure~\ref{fig14} ($t = 3.1 \times 10^7$
years) and Figure~\ref{fig15} presents a more dramatic difference in
the morphology of the PVD and the corresponding dependence on the
viewing angle.  From this particular viewing angle (labeled 2), the
nuclear emission is dominated by a linear feature intersecting the
origin, surrounded by a large elliptical emission that mainly arises
from the structure of the developing nuclear ring.  The emission along
any radial offset from the origin spans a large range in velocity
since different velocity components from physically different parts of
the inflow contributes significantly to the emission in the PVD.  For
example at an offset distance of +1 kpc (see Figure~\ref{fig15}), more
than half of the emission is seen in the redshifted part of the PVD
(positive velocity), which is expected if the gas particles follow
normal rotation.  However, it is evident that a significant fraction
of emission at +1 kpc populates the blueshifted half of the PVD
(negative velocity) -- the so-called ``forbidden velocity quadrant''
-- because of substantial contribution from the inflow from the
opposite side of the galaxy along this line of sight.  The viewing
angle from position 2 is one such example that shows substantial
emission in the forbidden velocity quadrant, although emission in the
forbidden velocity quadrant is almost always present to some degree as
seen in Figure~\ref{fig14}.  This a powerful way to identify transient
and anomalous kinematics in observations with adequate spatial and
velocity resolution.  However, the information contained in the PVD is
not enough to distinguish between infall and outflow, and a robust,
self-consistent interpretation requires an a priori knowledge of the
dynamical state of the source.

The Taffy Galaxy is an example of two disk systems that have recently
undergone a penetrating collision, evidence of which comes from the
long stretching radio continuum bridges seen between the two galaxies
\citep{cond93,cond02}.  The PVD shown in Figure~\ref{fig16} presents
substantial emission in the forbidden velocity quadrant (dotted
circle) which is kinematically and spatially distinct from the rest of
the galaxy.  The above analysis suggests that one possible implication
of such distinct emission in the forbidden velocity quadrant is inflow
(or outflow) caused by the immense head-on collision that occurred
$\sim 10^7$ years ago \citep{cond93}.

\clearpage

\begin{figure*}
\epsscale{1}
\caption{The evolution of the PVDs are shown in rows, where each is separated 
into three figures according to the adopted density cutoff.  The PVDs 
constructed using all of the particles are shown in $left$, 
$n_{\rm gas} \geq 1.0$ are shown in $middle$ and $n_{\rm gas} \geq 10$ are 
shown in $right$.  The spatial and velocity pixel resolution is set at 
$\sim 216$pc and $\sim 26$ km/s respectively to best represent the 
resolution achievable with current interferometric instruments.}
\label{fig14}
\end{figure*}

\begin{figure*}
\epsscale{1}
\caption{The PVD of the simulated galaxy viewed from position 1 ($left$) 
and position 2 ($right$) in Figure 2.  Compare this to the first row in 
Figure 14. }
\label{fig15}
\end{figure*}

\begin{figure}
\epsscale{1}
\caption{The PVD of UGC~12915 (the ``Taffy Galaxy'').  The slit is placed 
at the major axis of the edge on galaxy centered at the K-band peak.  
Substantial amount of emission is seen in the ``forbidden velocity quadrant'' 
(dotted circle) whose  kinematics is clearly distinct from the rest of the 
galaxy.}
\label{fig16}
\end{figure}

\clearpage

\subsection{Rotation Curve Derived from the PVD and the Derivation of M$_{dyn}$}

A well sampled rotation curve allows an indirect measurement of the
mass distribution of galaxies.  For example, in many rotationally
supported normal spiral galaxies, the observational evidence of flat
CO and \ion{H}{1} rotation curves indicate the ubiquitous presence of
massive dark matter halos \citep[see][for a review]{sofue01}.
Spectroscopic observations of atomic and molecular emission lines have
been the most common approach to derive the rotation curves of low
velocity dispersion, rotationally supported disks \citep{sofue01}.  In
colliding systems, however, tidal interactions can drive strong
non-circular motion in gas which can in turn significantly impact the
interpretation of the observed PVD and the derivation of a rotation
curve.  Not all colliding systems are readily identified as such based
on optical appearance alone \citep[e.g. M81, ][]{yun94}, and tidally
driven non-circular motion may play an important role in many apparent
isolated galaxies. Therefore, it is important to understand the
accuracy, reliability and the limitation of the rotation curves
derived from a simulated disk-disk collision first, in order to gain
confidence in the interpretation of the observational results.
Several rotation velocity fitting techniques are commonly used in the
literature (see \citet[][]{sofue01} for a review) such as the (1)
terminal velocity method \citep{sofue96}, (2) fitting the lowest
emission corrected by the instrumental resolution and the ISM velocity
dispersion (the envelope method) \citep[e.g.][]{sancisi79}, (3)
fitting a Gaussian to the line profile, or (4) tracing the peak
intensity.  Line of sight projection can complicate the use of many of
these methods in highly inclined galaxies \citep{uson03}, but the peak
intensity method works reasonably well for optical spectroscopic
observations of low inclination galaxies with good S/N \citep{rubin80,
mathewson96}.  A Gaussian can represent the line profile in similarly
low inclination galaxies, but the approximation fails again for
edge-on galaxies when substantial beam smearing occurs.  A commonly
used technique to avoid this problem is to construct a PVD of a galaxy
along the kinematic major axis, and fit the edges of the emission
contours using techniques such as the terminal velocity or envelope
methods.  In brief, the rotation velocity from the terminal velocity
method is defined as
\begin{equation}
V_{\rm r} = V_{\rm t} - V_{\rm sys} - \sqrt{\sigma^2_{\rm obs} + \sigma^2_{\rm ISM}}
\end{equation}
where $V_{\rm sys}$ is the systemic velocity, $\sigma_{\rm obs}$ is
the velocity resolution of the data, and $\sigma_{\rm ISM}$ is the
velocity dispersion of the interstellar medium.  The terminal
velocity, $V_{\rm t}$, is defined as the velocity at $I_{\rm t} =
\sqrt{(\eta I_{\rm max})^2 + I_{\rm min}^2}$, where $I_{\rm max}$ and
$I_{\rm min}$ are the peak and lowest emission respectively, and the
suggested value of $\eta = 0.2$ is used to represent the emission
intensity at the 20\% level \citep{sofue01}.  Experiments show that
this method works well for isolated galaxies of various inclinations,
except for the inner most regions of the galaxies where the angular
resolution limits the goodness of the fit \citep{sofue96}.  The
envelope method is a less sophisticated technique, correcting only for
the velocity resolution and the ISM turbulence and not weighting the
peak intensity.  Strictly speaking, the application of any of these
methods to simulations is subject to uncertainties since the PVD in
simulations can only trace the gas density and not the emission
intensity.  Since the emission intensity depends on the optical depth,
excitation conditions, and the beam filling factor --- all of which
varies from one tracer to another --- we assume here for simplicity
that the emission intensity is proportional to the gas column density
(i.e. the optically thin case).  For all of the derived rotation
curves in what follows, the turbulence term ($\sigma^2_{\rm ISM}$) is
neglected for consistency since the contribution is small ($\sim 10$
km/s) and the exact value varies from source to source.

\begin{figure*}
\epsscale{1}
\plotone{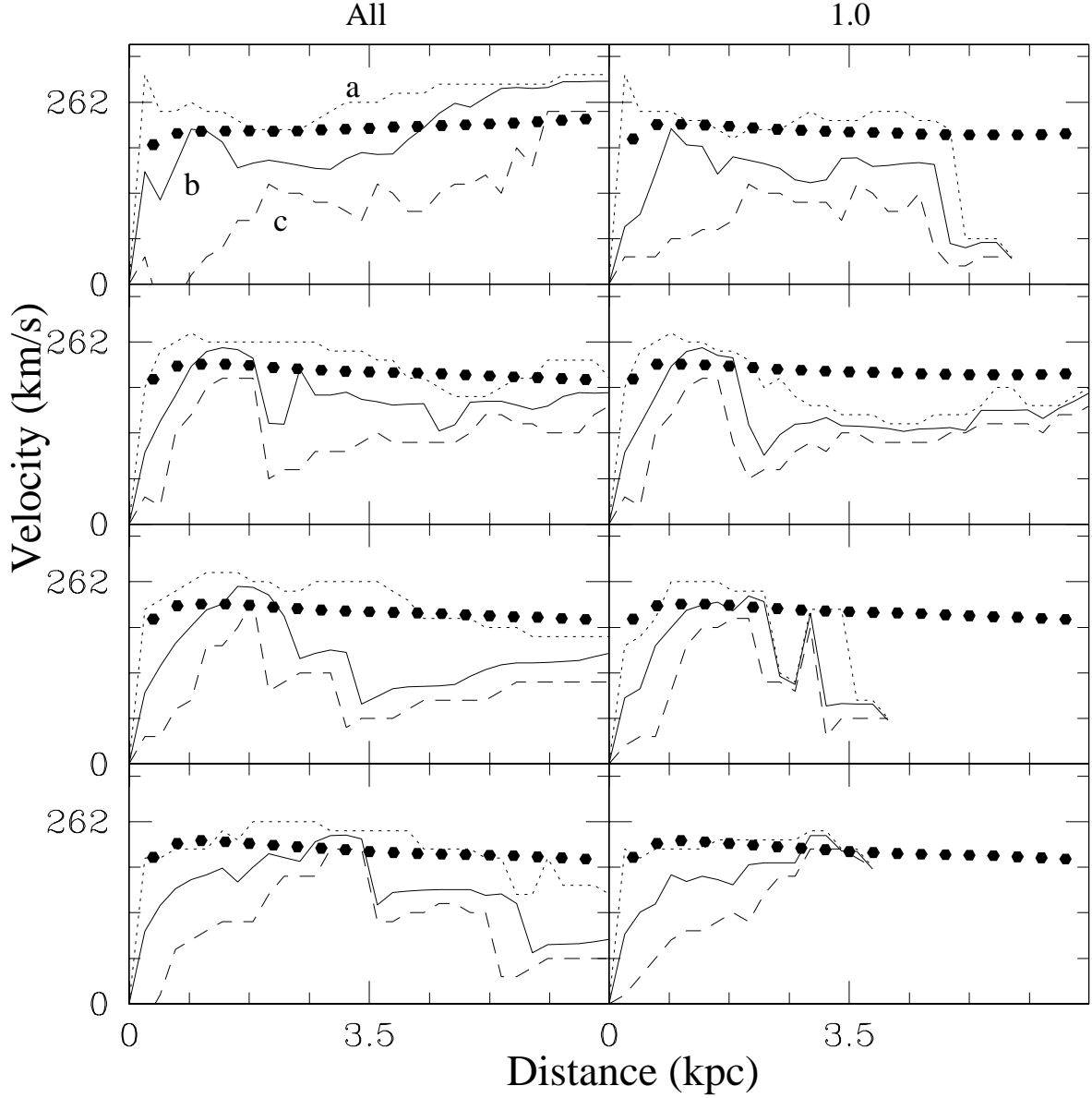}
\caption{The envelope method (\textit{dotted line} labeled a), terminal 
velocity method (\textit{solid line} labeled b) and maximum intensity fitting 
(\textit{dashed line} labeled c) are used to derive the rotation curves to 
the evolution of the PVD shown in Figure 14, with time advancing from top 
to bottom.  The solid dots represent the true rotation curves derived by 
using $V_{\rm rot} = (GM/R)^{1/2}$.  The rotation curve derived to the PVD 
with no density cutoff (``all'' in Figure 12) is shown in $left$ while  
$right$ shows the rotation curve to the PVD when the emission is restricted 
to $n_{\rm gas} \geq 1.0$.}
\label{fig17}
\end{figure*}

The rotation curves during the massive inflow period are shown in
Figure~\ref{fig17}.  Since the higher density gas ($n_{\rm gas} \geq
10$ cm$^{-3}$) is clumpy and only abundant in the inner few kpc of the
galaxy, the rotation curve fitting is done only on the first two
columns of Figure~\ref{fig14} (``All'' and $n_{\rm gas} \geq 1.0$
cm$^{-3}$).  It is clear from Figure~\ref{fig17} that simply tracing
the maximum intensity does not offer a reliable fit of the rotation
curve at any epoch.  The largest discrepancy is found in the outer
disk because the dense gas does not fill the disk uniformly and the
PVD is not fully sampled.  The terminal velocity method and 
the maximum intensity fitting both   significantly underestimate the
true rotation in the outer disks.  The
envelope method, on the other hand, appears to offer a reasonably good
fit to the true rotation curves at all scales up to 7 kpc, possibly
because the envelope kinematics are dominated by the limb-brightened
tangent points.

\clearpage

\clearpage

\begin{figure*}
\epsscale{1}
\plottwo{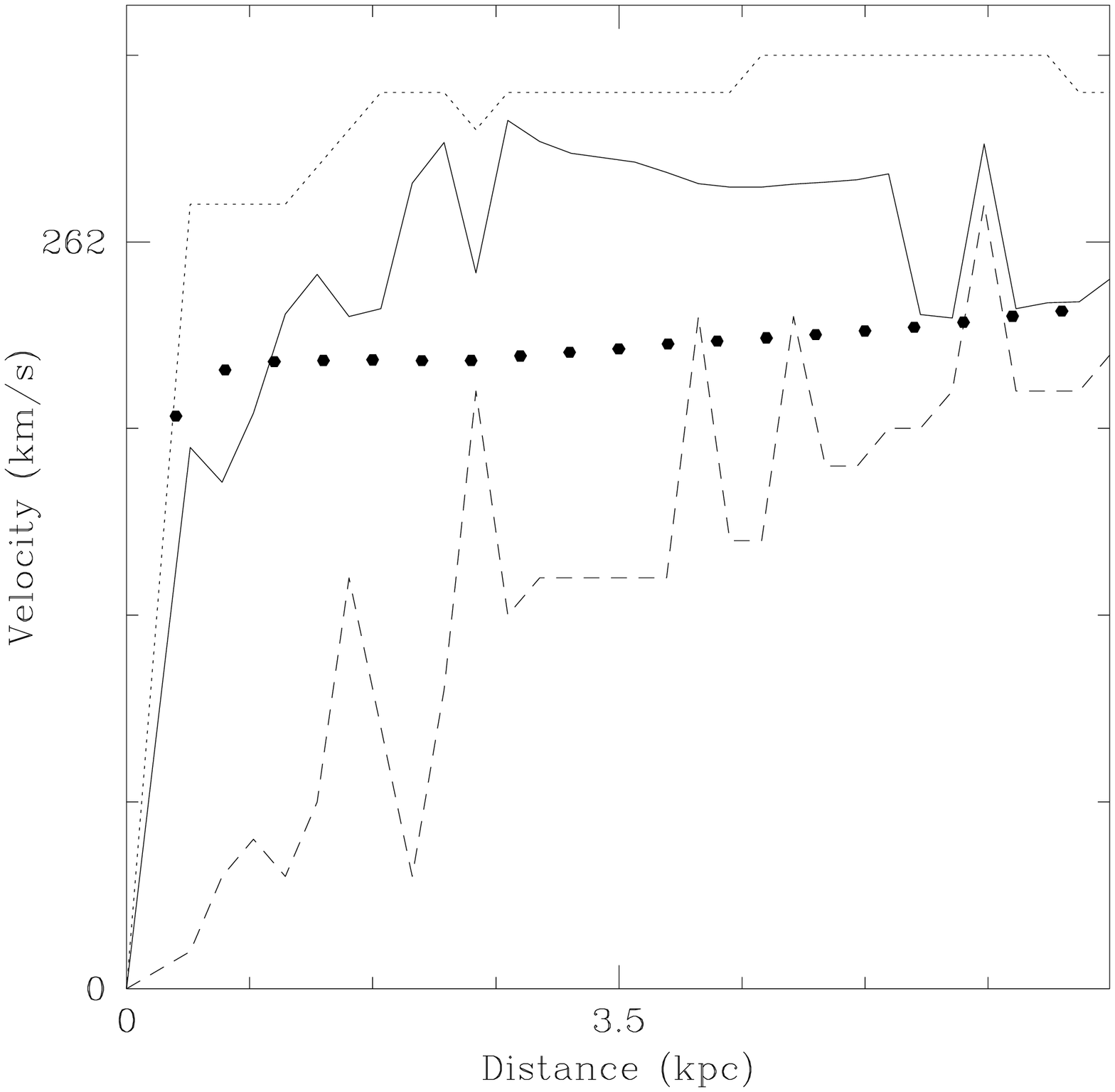}{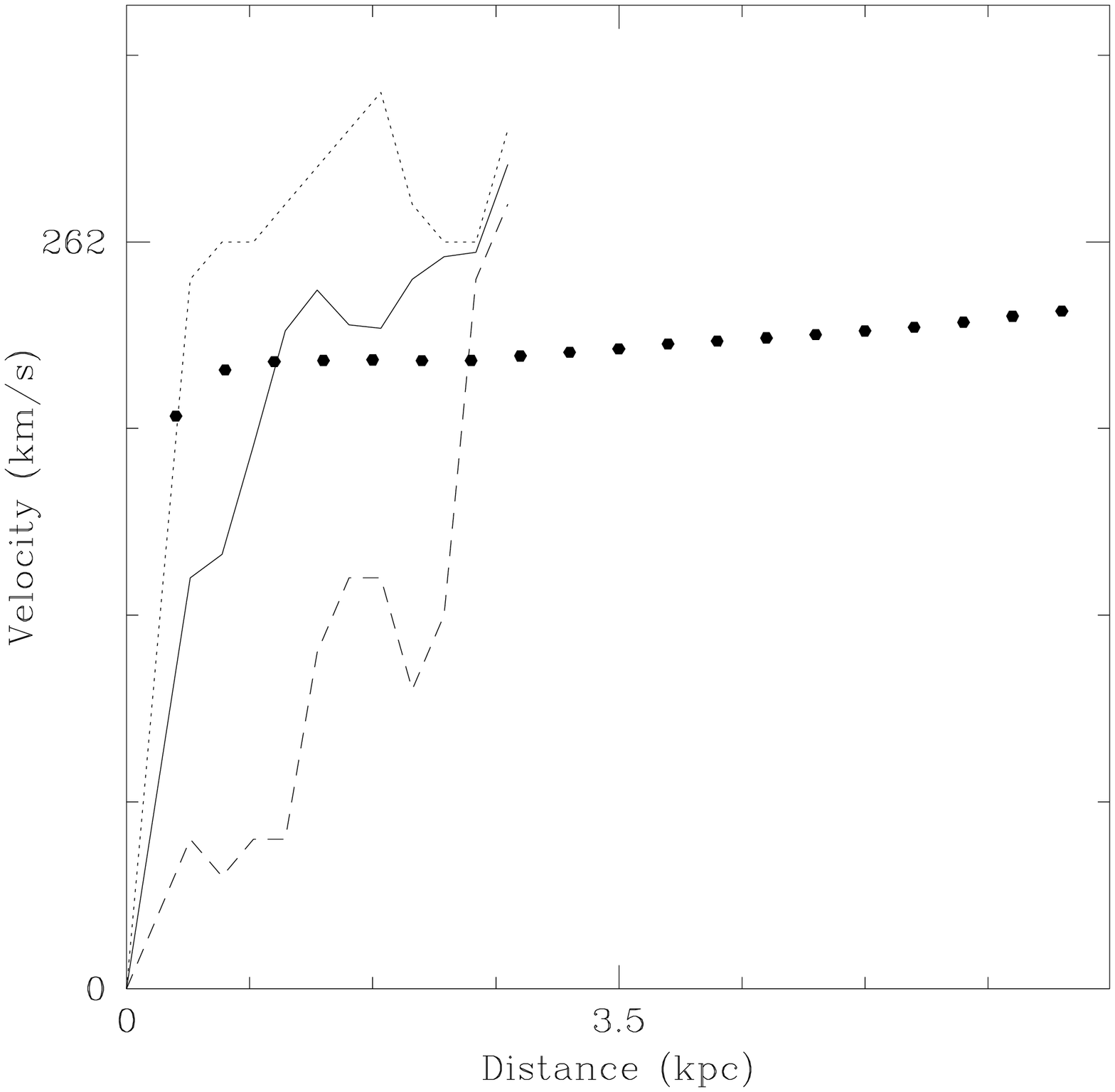}
\caption{Similar to Figure 17, but the PVD shown in Figure 15 (viewed 
from position 2) is used to derive the rotation curves.}
\label{fig18}
\end{figure*}

Figure~\ref{fig18} presents the rotation curve derived using the PVD
constructed along angle 2. Since the shape of the PVDs depend strongly
on viewing geometry, the resulting rotation curves also show
significant variations depending on the adopted line of sight toward
the inner disk.  In concordance with such expectation, the envelope
method now significantly overestimates the rotation in the inner few
kpc, but the rotation in the outer disk is consistent with that
derived along angle 1 (Figure~\ref{fig17}).  Since the inflow
predominantly affects the shape of the PVD inside the effective bar
potential (which occurs only in the inner few kpc), the dependence on
the viewing angle effectively diminishes at large radii.

\clearpage

\clearpage

Table 1 summarizes the dynamical mass ($M_{\rm dyn} = V^2_{rot}R/G$)
calculated at different radii using the rotation curve derived from
the terminal velocity method, and Table 2 presents the results
obtained using the envelope method.  Assuming a spherical mass
distribution for a flattened potential introduces a $\sim 10 - 30$\%
uncertainty in the computed dynamical mass \citep{binney87}, with the
exact correction factor depending on the details of the true mass
distribution.  Table~1 also compares the dynamical mass with the true
mass obtained directly from the simulation by summing, gas, stars, and
dark matter particles enclosed in a sphere defined by $R$.  On
average, the derived dynamical mass underestimates the true mass by
$20 - 40$\%, and sometimes as large as 80\% ($t = 9.3$ and $12 \times
10^7$ years).  The dynamical mass in the inner disk is underestimated
by $20 - 30$\% even before the collision ($t = -19 \times 10^7$ years)
when the disk kinematics are mostly dominated by rotation and not by
inflow.  The results presented in Table~1 are merely a reiteration of
the previous argument that the terminal velocity method often results
in inaccuracies in deriving the true rotation curve.  In contrast, it
is readily seen that the results from the envelope method presented in
Table~2 are more consistent with the true mass.  A $10 - 30$\% error
can easily be accounted for by the spherical mass approximation, thus
the results here imply that the envelope method can predict the
enclosed mass to within $20 - 40$\% of the true mass.

\begin{figure*}
\epsscale{1}
\plottwo{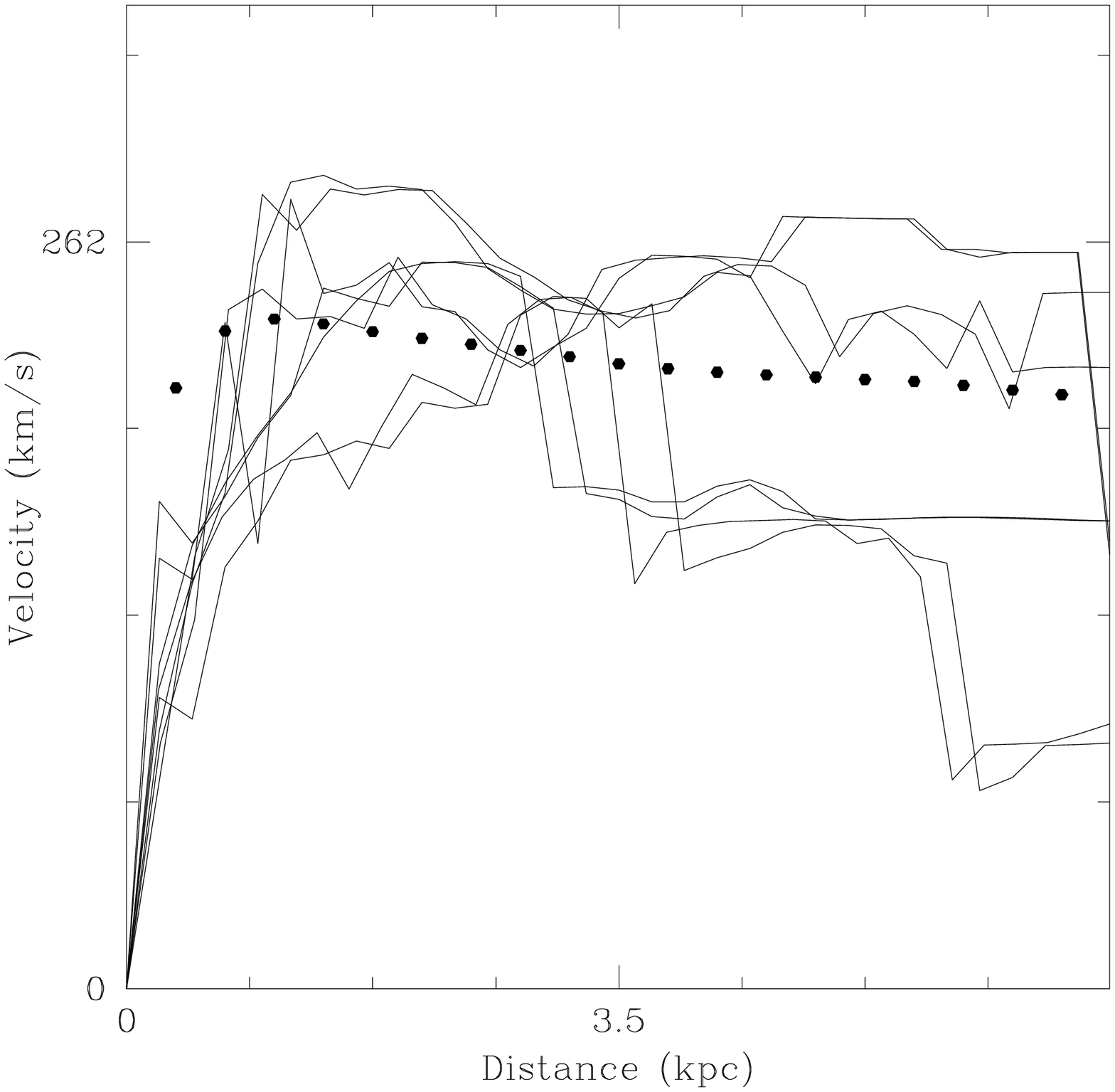}{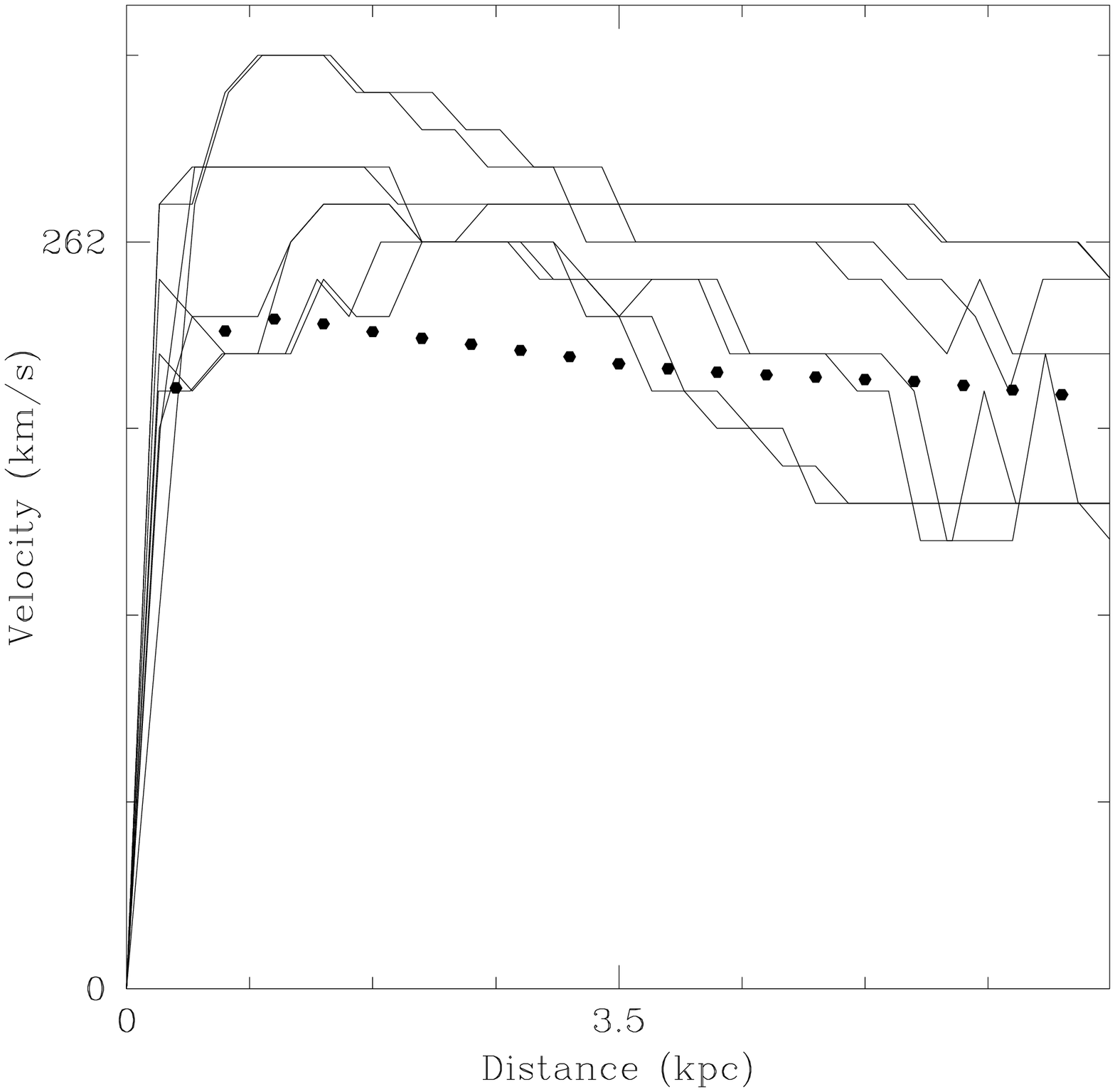}
\caption{Rotation curves derived for a disk with $i=90^{\circ}$ using the 
terminal velocity method employing various viewing angles in increments 
of $45^{\circ}$ in azimuth. Again, the solid dots represent the rotation 
curve derived using $V_{\rm rot} = (GM/R)^{1/2}$.}
\label{fig19}
\end{figure*}

Significant variations in the shape of the PVD and the resulting
rotation curve are seen when two viewing angles are compared.  To form
a complete set of systematically different viewing angles, a total of
7 rotation curves were derived by varying the viewing angle by
$45^{\circ}$ increments azimuthally around the galaxy at $t= 12 \times
10^7$ years.  The rotation curves presented in Figure~\ref{fig19} use
both the terminal velocity and the envelope methods for comparison.
The dotted points again represent the true rotation derived from the
spherically averaged mass distribution, which in principle should
reflect the average of all of the rotation curves derived from
different viewing angles. Using the terminal velocity method to
measure the rotation velocity in the inner few kpc works well, and the
dynamical mass is determined to within 10\% of the true mass.
However, significant deviations from the true rotation speed are
evident in the outer disk in about half of the viewing angles,
underestimating the true mass by 30\% on average.  In contrast, the
envelope method consistantly overestimates the true rotation curve and
the resulting dynamical mass.  Even when the tidal features dominate
the outer disk, on average the envelope method appears to trace the
true rotation better.

\clearpage

\begin{deluxetable}{crrrr}
\tablewidth{0pt}
\tablecaption{The Dynamical Mass (Terminal Velocity Method)\label{tabA2}}
\tablehead{
\colhead{time}& \multicolumn{4}{c}{$M_{\rm dyn}$($\times 10^{10} M_\odot$)}\\
\colhead{($\times 10^7$ yr)}&  \colhead{R = 1.8} &\colhead{R = 3.5}  &\colhead{R = 5.3} &\colhead{R = 7.0}
} 
\startdata
-19 & 1.1 (-35\%) & 3.0 (-22\%) & 5.9 (~-5\%) & 8.6 (~-2\%)\\
3.1 & 1.2 (-37\%) & 2.8 (-32\%) & 9.1  (~40\%) & 9.3 (~51\%) \\ 
6.2 & 2.3  (~11\%) & 2.5 (-35\%) & 3.9 (-31\%) & 5.8 (-24\%) \\ 
9.3 & 2.6 (~27\%) & 0.7  (-83\%) & 2.4 (-54\%) & 4.1 (-40\%)  \\
12  & 1.6 (-26\%) & 2.9 (-25\%) & 3.0 (-45\%) & 1.4 (-80\%) 
\enddata
\tablecomments{The percentage error from the true mass is shown in (), where a minus sign represents an underestimate from the true mass.}
\end{deluxetable}

\clearpage

\clearpage

\begin{deluxetable}{crrrr}
\tablewidth{0pt}
\tablecaption{The Dynamical Mass (Envelope Method)\label{tabA3}}
\tablehead{
\colhead{time}& \multicolumn{4}{c}{$M_{\rm dyn}$($\times 10^{10} M_\odot$)}\\
\colhead{($\times 10^7$ yr)}&  \colhead{R = 1.8} &\colhead{R = 3.5}  &\colhead{R = 5.3} &\colhead{R = 7.0}
} 
\startdata
-19 & 1.6 (-10\%)& 3.1 (-18\%) & 7.3 (~16\%)&  9.1 (~~3\%)\\
3.1 & 2.1 (~~9\%) & 5.6 (~37\%) & 10.2 (~57\%) & 14.8 (~71\%) \\
6.2 & 2.8 (~35\%) & 4.8 (~25\%) &4.5 (-20\%) &7.2 (~-6\%)\\
9.3 & 2.8 (~35\%) & 5.6 (~45\%) &4.7 (-14\%) &5.5 (-21\%)\\
12  & 2.5 (~18\%) & 5.0 (~30\%) &5.4 (~-3\%) &4.0 (-41\%) 
\enddata
\end{deluxetable}

\clearpage


\section{The use of Atomic and Molecular Observation to Determine the Merger Chronology}

The optical morphology of interacting systems conveys distinctive
information about the interaction history of the galaxy pairs.  The
shortcoming of this method is similar to the problem of classifying
galaxies into Hubble types (\S4) in that it is rather subjective and
potentially biased, making a more robust and systematic method
preferable.  A possible quantitative method is to use the compactness
parameter ($K$) in a statistically significant sample of colliding
galaxies as a tracer of dynamical age.  If Figure~\ref{fig13} holds
true universally, later stage interactions should have higher
compactness parameters.  Another possible approach is to characterize
the fraction of molecular gas to the total gas mass, since conversion
from atomic to molecular gas (and molecular gas to stars) is expected
to be more efficient in a colliding system.

The evolution of the molecular gas mass fraction ($M_{\rm H_2}/M_{\rm
H_2 + HI}$) during the simulated collision is presented in
Figure~\ref{fig20}.  A cutoff volume density of 1.0 cm$^{-3}$ is
adopted to roughly separate the gas that is dominated by atomic verses
molecular gas, which gives $M_{\rm H_2}/M_{\rm H_2 + HI} \sim 0.25$
initially.  Although slightly higher, this value is consistent with
the mean empirical ratio derived by \citet{casoli98} ($M_{\rm
H_2}/M_{\rm H_2 + HI} = 0.17$) using a distance-limited sample of 582
galaxies.  The sharp rise in the molecular fraction occurs immediately
after the collision and during the inflow period, followed by a
gradual decline while the system settles to the semi-equilibrium
state, as the inner nuclear gas slowly converts to stars.  The
molecular fraction peaks at $\sim 5 \times 10^8$ years when the two
galaxies finally coalesce, after which the molecular fraction plummets
due to the strong merger-induced starburst.

\subsection{Application to the Toomre Sequence}

\citet{toomre77} compiled a set of 11 nearby systems currently
experiencing collisions or mergers, and roughly ordered them according
to the completeness of the interaction purely based on optical
morphology (the so called ``Toomre Sequence'').  The Toomre Sequence
has since been studied in a number of ways including; IR
\citep[e.g.][]{joseph85}, \ion{H}{1} \citep[e.g.][]{hibbard96}, CO
\citep{yun01}, and numerical simulations
\citep[e.g.][]{barnes88,mihos93}.  Recently, \citet{laine03}
investigated the nuclear regions of the complete sample of the Toomre
Sequence using both the broad-band and narrow-band filters on the
HST/WFPC2.  One of their main goals was to search for trends in the
optical nuclear properties of the galaxies as a function of merger
age. No clear trends were found; while firm conclusions were difficult
to make due to the ubiquitous presence of dust in the nuclei, this
result may also point towards the need for further observational
diagnostics to address whether the Toomre Sequence represents the
correct canonical ordering of galaxy evolution.

\begin{figure}
\epsscale{1}
\plotone{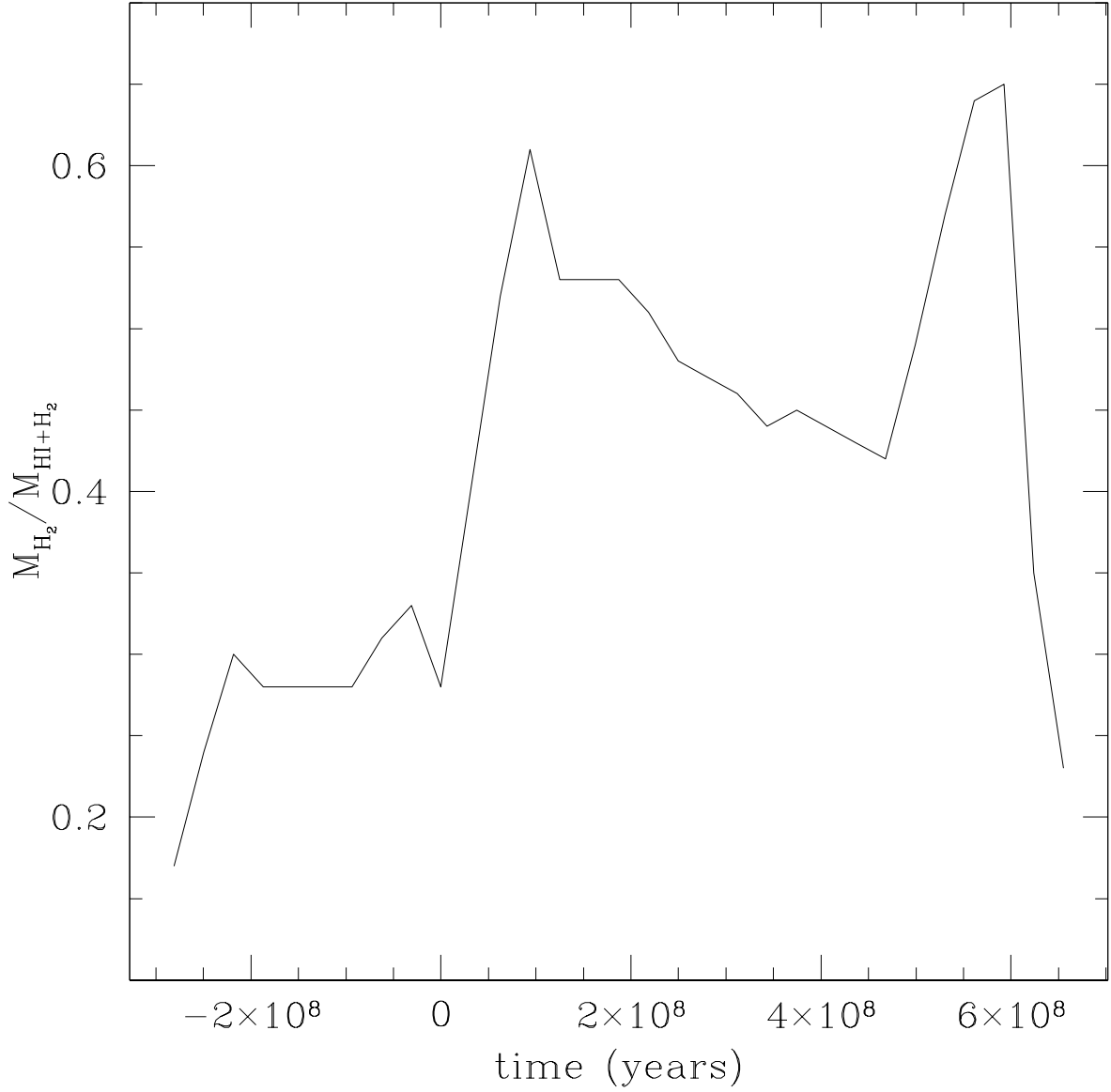}
\caption{Evolution of the estimated molecular gas  mass fraction 
($M_{\rm H_2}/M_{\rm H_2 + HI}$) throughout the course of the simulation. Molecular
gas is defined here to be gas above a volume density  of 1.0 cm$^{-3}$ (see text).}
\label{fig20}
\end{figure}

In order to study the gas properties along the Toomre Sequence, Table
3 shows the molecular fraction for six subsystems of the Sequence,
compiled from the extant literature. In general, the data does not
show an obvious trend in the molecular fraction as a function of the
proposed interaction sequence.  However, it may be possible to argue
that the first three systems (early stage mergers) have slightly
higher molecular fraction on average compared to the latter three
(late stage mergers) where the coalescence may have converted much of
the molecular gas to stars as was demonstrated in the simulation in
Figure~\ref{fig20}.  Direct comparison between observation and
simulation may prove difficult in the early stage mergers since the
sharp rise in molecular fraction (i.e. the massive inflow period) only
accounts for $10 - 20$\% of the total merger timescale which is then
followed by a gradual decline.  However, the possibility that the
molecular fraction is higher in the pre-merger phase is intriguing,
and a systematic survey of a larger sample should address the validity
of this analysis (Paper II).

\clearpage

\clearpage

\section{Summary}

We revisit numerical simulation results of \citet{mihos96} to investigate 
the response of gas in colliding systems and identify observational signature 
of inflow in the simulated data.  The key results found in this study are 
summarized below:

\begin{enumerate}

\item Response of the gas to the $m=2$ perturbation

Stars respond to the tidal interaction by forming both transient arms
and long lived $m=2$ bars, but the gas response is more transient,
flowing directly toward the central regions within $10^8$ years after
the initial collision. The rate of inflow declines when more than half
of the total gas supply reaches the inner few kpc, where the gas forms
a dense nuclear ring inside the stellar bar.  The orientation of the
ring suggests an orbital structure similar to those identified as $x_2$
orbits induced by the presence of an ILR.  The presence/absence of the
ILR may govern the strength and duration of the star formation
activity by regulating the gas inflow to the center of the galaxy as
seen, for example, in the bulgeless galaxy encounter.  Shocks are
identified using velocity discontinuities, which often spatially
coincides with the high density gas regions near the inflowing gas.  

\item Evolution of the structural parameters

The evolution of the asymmetry ($A$), concentration ($C$) and the new 
compactness ($K$) parameters were investigated for both the stars and gas.  
In contrast to the concentration ($C$) and compactness ($K$) parameters 
which work reasonably well, the asymmetry parameter ($A$) appears to be a 
poor tracer of the structural evolution of both stars and gas.  The strong 
evolution in $K$ may be used to infer the merger chronology of colliding 
systems and this will be further tested in paper II.

\item The use of the PVD as a diagnostic tool to infer inflow

Gas in non-circular kinematics can populate the ``forbidden velocity quadrant''
of the PVD, most distinctively seen when the line of sight intersects the 
infalling gas.  Three PVD fitting techniques were employed to derive the 
rotation curves, and the dynamical mass was estimated to test how accurately 
the true mass can be recovered.  The results show that the dynamical mass can 
be determined to within $20 - 40$\% of the true mass if the envelope method is 
adopted.  

\item Merger chronology from atomic/molecular gas observations 

The simulation results predict a marked increase in the molecular
fraction during the massive inflow period, but the application to real
systems may require additional observational constraints to properly
assess the exact chronology especially in the pre-mergers.  More
speculatively, the spatial distribution of atomic/molecular gas can be
used to convey information on the approximate merger age since the
galaxies have collided for the first time.  By combining molecular and
atomic gas observations in tidally interacting systems, the
determination of an approximate merger age may be possible.  This
hypothesis will be tested in our forthcoming paper (paper II).

\end{enumerate}

\acknowledgements

J.C.M. is supported by a Research Corporation Cottrell Scholarship and by the NSF 
through CAREER award AST 98-76143.

\clearpage

\begin{deluxetable}{lcc}
\tablewidth{0pt}
\tablecaption{Molecular Fraction in the Toomre Sequence\label{tabA1}}
\tablehead{
\colhead{System}&  \colhead{$M_{\rm H_2}/M_{\rm HI + \rm H_2}$} & \colhead{Age\tablenotemark{a}}
} 
\startdata
NGC~4038/9 \tablenotemark{b,c}  & 0.76 & E\\
NGC~4676  \tablenotemark{d}  & 0.61 & E\\
NGC~6621/2 \tablenotemark{e}  &  0.83 & I\\
NGC~520 \tablenotemark{d}  &  0.66 & IL\\
NGC~3921 \tablenotemark{d}  &  0.47 & L\\
NGC~7252 \tablenotemark{d}  &  0.64 & L \\
\enddata
\tablenotetext{a}{The interaction age assigned based on investigating the 
optical morphology by eye; E (\textit{early stage}), I 
(\textit{intermediate stage}), L (\textit{late stage}) }
\tablenotetext{b}{\citet{hibbard01}}
\tablenotetext{c}{\citet{gao01}}
\tablenotetext{d}{\citet{hibbard95}}
\tablenotetext{e}{Paper II}

\end{deluxetable}

\end{document}